\documentclass[10pt]{article}
\usepackage{blindtext}
\usepackage[T1]{fontenc}
\usepackage[utf8]{inputenc}
\usepackage{amssymb,amsfonts,amsmath,geometry,graphics,caption,amsmath, graphics, array, tabularx, graphicx, float,geometry, rotating,setspace, fancyhdr, mathtools, amssymb, caption, caption, latexsym, array, tabularx, subfig, multirow,epstopdf, booktabs,caption,fixltx2e,epstopdf,color,amsmath,bm,comment}
\usepackage{scalerel,stackengine}

\usepackage{float}
\usepackage{algorithm,algcompatible}
\algnewcommand\INPUT{\item[\textbf{Input:}]}%
\algnewcommand\OUTPUT{\item[\textbf{Output:}]}%

\usepackage{threeparttable}

\usepackage{adjustbox,lipsum}

\usepackage{setspace}
\usepackage{lineno}

\title{Detecting Imprinting and Maternal Effects Using Monte Carlo Expectation Maximization Algorithm}
\author{Pooya Aavani, Alexandre Trindade, Fangyuan Zhang}
\date{}

\usepackage{natbib}

\begin{document}

\maketitle
\author{}
\begin{abstract}
Numerous statistical methods have been developed to explore genomic imprinting and maternal effects,
which are causes of parent-of-origin patterns in complex human diseases. However, most of them either
only model one of these two confounded epigenetic effects, or make strong yet unrealistic assumptions
about the population to avoid over- parameterization. A recent partial likelihood method (LIME)
can identify both epigenetic effects based on case-control family data without those assumptions.
Theoretical and empirical studies have shown its validity and robustness. However, because LIME
obtains parameter estimation by maximizing partial likelihood, it is interesting to compare its efficiency
with full likelihood maximizer. To overcome the difficulty in over-parameterization when using full
likelihood, in this study we propose a Monte Carlo Expectation Maximization (MCEM) method to
detect imprinting and maternal effects jointly. Those unknown mating type probabilities, the nuisance
parameters, can be considered as latent variables in EM algorithm. Monte Carlo samples are used to
numerically approximate the expectation function that cannot be solved algebraically. Our simulation
results show that though this MCEM algorithm takes longer computational time, and can give higher bias in some simulations compared to LIME, it can generally detect both epigenetic effects with higher power and smaller standard error which demonstrates that it can be a good complement of LIME method.

\end{abstract}


\textbf{Keywords}: Imprinting, Maternal effect, Monte Carlo Expectation Maximization, Full likelihood, Case-control family data

\section*{Introduction}

Through the process of reproduction, descendants inherit the genetic information from their parents. Many hypotheses proposed to explain the mechanism of inheritance but only two them are considered to be compelling. Mendelian inheritance explains heredity to be consistent with the transmission of DNA such that the occasional changes in DNA sequences can happen by mutation or recombination during the process of transmission \citep{boffelli2012epigenetic}. But in chromosome, there are some materials that are distinct from the DNA and capable of being transmitted alongside of it. These materials can change the expression of the genes that is independent of mechanisms such as mutation and therefore can affect the phenotype during the cell division. The non-Mendelian type of inheritance in such cases is called epigenetic inheritance. For example, epigenetic inheritance can occur through reversible modifications of the genetic materials such as DNA methylation \citep{franklin2010epigenetic}, which results inactivation of the gene expression from addition of methyl group to the specific residue in DNA \citep{read2018human}.

Genomic imprinting and maternal effects are two important confounding epigenetic effects that can both lead to parent-of-origin pattern\citep{hager2008maternal}. Genomic imprinting  is the phenomenon in which genes in offspring are differentially expressed due to different parental origins \citep{barlow2011genomic}. Maternal effect refers to the phenomenon that the child's phenotype depends on the mother's genotype, rather than the child's own genotype \citep{wolf2009maternal}. Several diseases are found to be related to imprinting effect such as Angelman syndrome and Prader-Willi syndrome that distrust the normal development of the child  \cite{heksch2017review}. Maternal effect can also be responsible for diseases such as obesity, hypertension, coronary artery disease, and non-insulin-dependent diabetes \cite{lillycrop2011effect}.

In the past decades, genome-wide association studies (GWAS) have identified many genetic variants associated with human complex traits, and tremendously increased our understanding about architectural genetics of those traits. These genetic variants, however, can only explain small proportion of variation in the traits, which leads to an effort of finding the "missing heritability" that includes both imprinting and maternal effects \citep{manolio2009finding}.

Both nonparametric and parametric methods have been suggested to study imprinting and maternal effects. Most nonparametric methods are used for detecting only imprinting effect by assuming no maternal effect\citep{weinberg1999methods,zhou2009detection}. These methods suffer from inflated false positive or false negative rates under the existence of maternal effect\citep{lin2013assessing}. 
Some parametric methods can detect imprinting and maternal effects jointly based on case-parent/control-parent triads\citep{zhang2019imprinting}. To avoid over-parametrization problem, however, most of these methods, such as \citet{weinberg1998log}, rely on assumptions on mating probabilities in population, such as mating symmetry. A partial likelihood method for detecting imprinting and maternal effects (LIME) can avoid making such strong assumptions as it estimates parameters by maximizing partial likelihood that is free of nuisance parameters \citep{yang2013robust}. Both empirical and theoretical studies have shown LIME is valid, powerful, and robust \citep{zhang2016optimum}. On the other hand, because LIME obtains parameter estimation by maximizing partial likelihood, it is interesting to compare its efficiency with full likelihood maximizer. To avoid strong assumptions about mating type probabilities and overcome the difficulty in over-parameterization when using
full likelihood, we propose Monte Carlo Expectation Maximization (MCEM) method. EM algorithm is commonly used for getting full likelihood maximizer when there are missing data or latent variables involved. In this paper, we use nuisance parameters--mating type probabilities as latent variables, and apply EM algorithm to detect both imprinting and maternal effects jointly. Monte Carlo samples are used to numerically approximate the expectation function that cannot be solved algebraically.

In the following section, we introduce our statistical model and demonstrate the application of MCEM to obtain full likelihood maximizer. In section \ref{simulation}, we simulate data under eight disease models and 8 scenarios, apply both LIME method and MCEM algorithm to the simulation data, and compare their bias in section \ref{bias}, and type I error and power in section \ref{typeI}.

\section{Statistical Model}\label{Model}
\subsection{Full likelihood}
We consider a candidate marker with two alleles $M_1$ and $M_2$, where $M_1$ is the minor allele. We define $M$, $F$, $C$ as the number of variant allele(s) $M_1$ (0, 1 or 2)  carried by mother, father, and child, respectively. The variable $D$ denotes the disease status of child such that $1$ is considered as being affected and $0$ being unaffected. We use multiplicative relative risk model for the disease penetrance as follows:
\begin{eqnarray}
P(D=1|M,F,C)&=& \delta R_{1}^{I(C=1)} R_{2}^{I(C=2)} R_{im}^{I(C=1 \text{ }\text{\& from mother})}S_{1}^{I(M=1)} S_{2}^{I(M=2)}\label{parameter1}\\
P(D=0|M,F,C)&=& 1-P(D=1|M,F,C)\label{parameter2}
\end{eqnarray}

In the above equations, $\delta$ is the phenocopy rate of the disease in the population. $R_{1}$ and $R_{2}$ are relative risks due to $1$ and $2$ copies of the variant allele carried by offspring, respectively. $R_{im}$ is the relative risk due to the single copy of the variant allele being inherited from the mother. $S_1$ and $S_{2}$ are the maternal effect of $1$ and $2$ copies of the variant allele carried by mother, respectively. $I$ is defined as an indicator function. Our method is based on case-parent and control-parent triads with arbitrary number of additional siblings.

The joint probability of proband's disease status and family genotype combinations, $P(D,M,F,C)$, can be written as follows:
\begin{equation}\label{dis_status}
P(D,M,F,C)= P(M,F)P(C|M,F)P(D|M,F,C)
\end{equation}

In Equation (\ref{dis_status}), $P(M=m,F=f)=\mu_{mf}$, denotes the probability (proportion) of parental pairs in which  mother and father carries $m$ and $f$ copies of the variant allele, respectively. In general, the mating type probabilities are missing or very hard to assess. $P(C=c|M=m,F=f)$ is the probability that child inherits $c$ variant allele(s) when mother has $m$ variant allele(s) and father has $f$ variant allele(s). This probability follows Mendel's law of segregation. There are $15$ possible genotype combinations of triads. Table \ref{tab:addlabel} demonstrates the joint probability of disease status and familial genotype combination.

\begin{table}[!htbp]
\centering
\caption{Joint probabilities of child's disease status and triad genotypes}
  \begin{threeparttable}[t]
  \centering
       \begin{tabular}{lr l r l r l r l r}
    \toprule
    Type & $M$     & $F$ & $C$  & $P(D=1,M,F,C)$ & $P(D=0,M,F,C)$\\
    \midrule
    1  &  $0$ & $0$ & $0$ & $\mu_{00}.1.\delta$   & $\mu_{00}.1.[1-\delta]$ \\
    2  &  $0$ & $1$ & $0$ & $\mu_{01}.\dfrac{1}{2}.\delta$   & $\mu_{01}.\dfrac{1}{2}.[1-\delta]$ \\
    3  &  $0$ & $1$ & $1$ & $\mu_{01}.\dfrac{1}{2}.\delta R_1$   & $\mu_{01}.\dfrac{1}{2}.[1-\delta R_1]$ \\
    4  &  $0$ & $2$ & $1$ & $\mu_{02}.1.\delta R_1$   & $\mu_{02}.1.[1-\delta R_1]$ \\
    5  &  $1$ & $0$ & $0$ & $\mu_{10}.\dfrac{1}{2}.\delta S_1$   & $\mu_{10}.\dfrac{1}{2}.[1-\delta S_1]$ \\
    6  &  $1$ & $0$ & $1$ & $\mu_{10}.\dfrac{1}{2}.\delta S_1 R_1 R_{im}$   & $\mu_{10}.\dfrac{1}{2}.[1-\delta S_1R_1 R_{im}]$ \\
    7  &  $1$ & $1$ & $0$ & $\mu_{11}.\dfrac{1}{4}.\delta S_1$   & $\mu_{11}.\dfrac{1}{4}.[1-\delta S_1]$ \\
    8  &  $1$ & $1$ & $1$ & $\mu_{11}.\dfrac{1}{4}.\delta S_1R_1(1+R_{im})$   & $\mu_{11}.\dfrac{1}{4}.[2-\delta S_1R_1(1+R_{im})]$ \\
    9  &  $1$ & $1$ & $2$ & $\mu_{11}.\dfrac{1}{4}.\delta S_1R_2$   & $\mu_{11}.\dfrac{1}{4}.[1-\delta S_1R_2]$ \\
    10  &  $1$ & $2$ & $1$ & $\mu_{12}.\dfrac{1}{2}.\delta S_1R_1$   & $\mu_{12}.\dfrac{1}{2}.[1-\delta S_1R_1]$ \\
    11  &  $1$ & $2$ & $2$ & $\mu_{12}.\dfrac{1}{2}.\delta S_1R_2$   & $\mu_{12}.\dfrac{1}{2}.[1-\delta S_1R_2]$ \\
    12  &  $2$ & $0$ & $1$ & $\mu_{20}.1.\delta S_2R_1R_{im}$   & $\mu_{20}.1.[1-\delta S_2R_1R_{im}]$ \\
    13  &  $2$ & $1$ & $1$ & $\mu_{21}.\dfrac{1}{2}.\delta S_2R_1R_{im}$   & $\mu_{21}.\dfrac{1}{2}.[1-\delta S_2R_1R_{im}]$ \\
    14  &  $2$ & $1$ & $2$ & $\mu_{21}.\dfrac{1}{2}.\delta S_2R_2$   & $\mu_{21}.\dfrac{1}{2}.[1-\delta S_2R_2]$ \\
    15  &  $2$ & $2$ & $2$ & $\mu_{22}.1.\delta S_2R_2$   & $\mu_{22}.1.[1-\delta S_2R_2]$ \\
     \bottomrule
  \end{tabular}
    \end{threeparttable}%
  \label{tab:addlabel}%
\end{table}%

$n^{1}_{mfc}, n^{0}_{mfc}$ are the numbers of case-parent and control-parent triads falling into $mfc$ genotype combinations, respectively. Similarly, we define $sn^{1}_{mfc}$ and $sn^{0}_{mfc}$ for sibling-parent triads.

Denote $Y=(n^{1}_{mfc}, n^{0}_{mfc},sn^{1}_{mfc},sn^{0}_{mfc})$ , $m=0, 1, 2$, $f=0, 1, 2$, $c=0, 1, 2$ as observed data. The full likelihood for case-parent/control-parent triads with arbitrary additional siblings is 
\begin{eqnarray}
f(Y|\bm{\mu};\bm{\theta}) &=&  \displaystyle \prod_{(m,f,c)} P(M=m,F=f,C=c|D=1)^{n_{mfc}^1} \times P(M=m,F=f,C=c|D=0)^{n_{mfc}^0}\notag\\
    &\times& P(D=1|M=m,F=f,C=c)^{sn_{mfc}^1}\times P(D=0|M=m,F=f,C=c)^{sn_{mfc}^0},\label{full-like}
\end{eqnarray} 
where $\bm{\mu}=(\mu_{00},\mu_{01},\mu_{02},\mu_{10},\mu_{11},\mu_{12},\mu_{20},\mu_{21},\mu_{22})$. As we have 15 unknown parameters, it is hard to get full likelihood maximizer directly. This is the reason why many methods need to make strong assumptions about $\bm{\mu}$ to reduce the number of free parameters.

To avoid strong assumptions about nuisance parameters, we use expectation maximization algorithm to get full likelihood maximizer. The nuisance parameters $(\mu_{00},\mu_{01},\mu_{02},\mu_{10},\mu_{11},\mu_{12},\mu_{20},\mu_{21},\mu_{22})$ are used as latent variables, and denoted as $\bm{Z}=(Z_{00},Z_{01},Z_{02},Z_{10},Z_{11},Z_{12},Z_{20},Z_{21},Z_{22})$.

\subsection{CEMC algorithm}

As $\sum \mu_{mf}=1$, we assume $\bm{Z}$ follows Dirichlet distribution with following probability density function:
\begin{equation}\label{Z-dist}
g(\bm{Z};\alpha)= \frac{1}{B(\bm{\alpha})} \prod_{m=0}^{2}\prod_{f=0}^{2} Z_{mf}^{(\alpha_{mf}-1)}
\end{equation}
where unknown parameter $\alpha$ is defined as follows:
\begin{equation}
\bm{\alpha}=(\alpha_{00},\alpha_{01},\alpha_{02},\alpha_{10},\alpha_{11},\alpha_{12},\alpha_{20},\alpha_{21},\alpha_{22}),
\end{equation}
and
\begin{equation}
B(\bm{\alpha})= \dfrac{\prod\limits_{m=0}^{2}\prod\limits_{f=0}^{2} \Gamma(\alpha_{mf})}{\Gamma(\sum\limits_{m=0}^{2}\sum\limits_{f=0}^{2}\alpha_{mf})}
\end{equation}

Denote $\bm{\psi}=(\bm{\theta},\bm{\alpha})$. The conditional probability distribution for $\bm{Y}$ given $\bm{Z}$, $P(\bm{Y}|\bm{Z}=\bm{\mu}; \bm{\theta})$, is the same as the original full likelihood $P(\bm{Y};\bm{\mu},\bm{\theta})$, except that the mating type probabilities are now viewed as latent random variables, rather than fixed parameters.

Log of the likelihood for complete data $(\bm{Y}, \bm{Z})$ is
\begin{equation}
l_{c}(\bm{\psi}; \bm{Y}, \bm{Z})= \log{f(\bm{Y}|\bm{Z};\bm{\theta})}+ \log{g(\bm{Z};\bm{\alpha})} \label{max-li1}
\end{equation}.

According to EM algorithm, we will calculate $Q(\bm{\psi}, \bm{\psi^{(t)}})=E_{\bm{\psi^{(t)}}}(l_{c}(\bm{\psi}; \bm{Y}, \bm{Z})|Y)$, and then update parameter estimation by maximizing $Q(\bm{\psi}, \bm{\psi^{(t)}})$ with respective to $\bm{\psi}$ until the estimation converges. However, as $Q(\bm{\psi}, \bm{\psi^{(t)}})$ cannot be solved explicitly, we use a Monte Carlo version $Q_{MC}(\bm{\psi}, \bm{\psi^{(t)}})$   to estimate it instead. The details are as follows:
\\
\textbf{E-Step}: 

We first calculate $Q(\bm{\psi};\bm{\psi}^{(t)})$ as
\begin{eqnarray}
Q(\bm{\psi};\bm{\psi^{(t)}})= E_{\bm{\psi^{(t)}}}\{l_{c}(\bm{\psi}; \bm{Y}, \bm{Z})|\bm{Y}\}=E_{\bm{\psi^{(t)}}}\{\log {f(\bm{Y}|\bm{Z};\bm{\theta})|\bm{Y}}\}+E_{\bm{\psi^{(t)}}}\{\log{g(\bm{Z};\bm{\alpha})|\bm{Y}}\}, \label{E-step2}
\end{eqnarray}
where 
\begin{eqnarray}
\log{f(\bm{Y}|\bm{Z};\bm{\theta})}= & \displaystyle \sum_{m,f,c}& \left\{ n_{mfc}^1 \log{P(M,F,C|D=1)}\right\}+ \sum_{m,f,c}\left\{n_{mfc}^0 \log{P(M,F,C|D=1)}      \right\} \label{derive1}\notag\\
+& \displaystyle \sum_{m,f,c}& \left\{ sn_{mfc}^1 \log{P(D=1|M,F,C) }\right\}+ \sum_{m,f,c}\left\{sn_{mfc}^0 \log{P(D=0|M,F,C) },     \right\}\label{derive11}\notag\\
\end{eqnarray}
and
\begin{eqnarray}
\log\left\{ g(\bm{Z};\bm{\alpha})\right\}  &=& -\log{B(\bm{\alpha})} +(\alpha_{mf}-1) \sum_{m=0}^{2}\sum_{f=0}^{2} \log{Z_{mf}}. \label{deriv3}
\end{eqnarray}
Therefore, the first term in $Q(\bm{\psi};\bm{\psi}^{(t)})$ can be expanded as
\begin{eqnarray}
E_{\bm{\psi^{(t)}}}\{\log {f(\bm{Y}|\bm{Z};\bm{\theta})|\bm{Y}}\}
&=& \int{\sum_{m,f,c} \left\{ n_{mfc}^1 \log{P(M,F,C|D=1)}\right\}} f_{\bm{\psi^{(t)}}}(\bm{Z}|\bm{Y})d\bm{Z}\label{int1}\notag\\
 &+& \int{\sum_{m,f,c} \left\{ n_{mfc}^0 \log{P(M,F,C|D=1)}\right\}}f_{\bm{\psi^{(t)}}}(\bm{Z}|\bm{Y})d\bm{Z}\label{int2}\notag\\
 &+& \int{\sum_{m,f,c} \left\{ sn_{mfc}^1 \log{P(D=1|M,F,C)}\right\}} f_{\bm{\psi^{(t)}}}(\bm{Z}|\bm{Y})d\bm{Z}\label{int3}\notag\\
 &+& \int{\sum_{m,f,c} \left\{ sn_{mfc}^0 \log{P(D=0|M,F,C)}\right\}}f_{\bm{\psi^{(t)}}}(\bm{Z}|\bm{Y})d\bm{Z}\notag\\
 \label{int4}
\end{eqnarray}
and the second term can be expanded as 
\begin{eqnarray}
E_{\bm{\psi^{(t)}}}\left(\log\left\{ \log{g(\bm{Z};\bm{\alpha})}\right\}  \big{|}\bm{Y}  \right)
&=& \int{\left( \left\{-\log{B(\bm{\alpha})} +(\alpha_{mf}-1) \sum_{m=0}^{2}\sum_{f=0}^{2} \log{Z_{mf}}\right\}\right)
 f_{\bm{\psi^{(t)}}}(\bm{Z}|\bm{Y})d\bm{Z}  }\notag\\
 \label{int5}
\end{eqnarray}.

\normalsize
As the integrals in Equations (\ref{int1}) and (\ref{int5}) are nonelementary and cannot be calculated explicitly, we use MCMC method to take samples $z^{(t,1)}, z^{(t,2)},..., z^{(t,10000)} $ from $f_{\bm{\psi^{(t)}}}(\bm{Z}|\bm{Y})$, where $z^{(t,i)}=(z^{(t,i)}_{00},z^{(t,i)}_{01},z^{(t,i)}_{02},z^{(t,i)}_{10},z^{(t,i)}_{11},z^{(t,i)}_{12},z^{(t,i)}_{20},z^{(t,i)}_{21},z^{(t,i)}_{22})$. Then we use these Monte Carlo samples to estimate the integrations.

$Q_{MC}(\bm{\psi}, \bm{\psi^{(t)}})=\widehat{E}_{\bm{\psi}^{(t)}}\{\log {f(\bm{Y}|\bm{Z};\bm{\theta})|\bm{Y}}\}+\widehat{E}_{\bm{\psi^{(t)}}}\{\log {g(\bm{Z};\bm{\alpha})|\bm{Y}}\}$, 
where 
\begin{eqnarray}
\widehat{E}_{\bm{\psi}^{(t)}}\{\log {f(\bm{Y}|\bm{Z};\bm{\theta})|\bm{Y}}\}&=& \dfrac{1}{10000} \sum_{i=1}^{10000} \sum_{m,f,c} \left\{ n_{mfc}^1\log{\left(\dfrac{ z_{mf}^{(t,i)}P(C|M,F)P(D=1|M,F,C)}{\sum\limits_{m,f,c}z_{mf}^{(t,i)}. P(C|M,F)P(D=1|M,F,C)}\right)} \right\}\notag\\
&+& \dfrac{1}{10000} \sum_{i=1}^{10000} \sum_{m,f,c} \left\{ n_{mfc}^0\log{\left(\dfrac{ z_{mf}^{(t,i)}P(C|M,F)P(D=0|M,F,C)}{\sum\limits_{m,f,c}z_{mf}^{(t,i)}. P(C|M,F)P(D=0|M,F,C)}\right)} \right\}\notag\\
&+& \dfrac{1}{10000} \sum_{i=1}^{10000}\sum_{m,f,c} \left\{ sn_{mfc}^1 P(D=1| M,F,C)   \right\} \notag\\
\label{Carlo3}
&+& \dfrac{1}{10000} \sum_{i=1}^{10000}\sum_{m,f,c} \left\{ sn_{mfc}^0 P(D=0| M,F,C)   \right\}, and \label{Carlo4}\\
\widehat{E}_{\bm{\psi^{(t)}}}\{\log {g(\bm{Z};\bm{\alpha})|\bm{Y}}\}&=& \dfrac{1}{10000} \sum_{i=1}^{10000}\left\{-\log{B(\bm{\alpha})} + \sum_{m=0}^{2}\sum_{f=0}^{2} (\alpha_{mf}-1) \log{z_{mf}^{(t,i)}}\right\}.\label{Carlo5}
\end{eqnarray}
A general method for generating samples from posterior distribution is Metropolis-Hastings (MH) algorithm \citep{Hastings,Givens05computationalstatistics}.

Simulate random vector $\bm{z}^{(i,t)}$ from the target distribution $k(\bm{Z})=f_{\bm{\psi^{(t)}}}(\bm{Z}|\bm{Y})$. Note that for $t=1$, we take $\bm{\alpha^{(1)}}_{mf}=\sum\limits_{c}\left(n_{mfc}^0\right) +1$ where $m=0,..,2$ and $f=0,..,2$. The reason for choosing such initial values for $\bm{\alpha^{(1)}}_{mf}$ is that mating type probabilities in control family are more similar as in general population.  $\bm{\theta}=(\delta,R_1,R_2,R_{im},S_1,S_2)=(0.0067,1,1,1,1,1)$. 

We define a proposal distribution $w(\bm{Z})=g(\bm{Z};\bm{\alpha}^{(t)})$. According to MH algorithm (see supplementary materials), in the first step we take sample $Z^{*}$ from $w(\bm{Z})$ distribution. Then we can calculate MH ratio as follow:
\begin{eqnarray}
R(\bm{Z}^{(t)},\bm{Z}^{*})&=& \dfrac{k(\bm{Z}^{*})w(\bm{Z}^{(t)})}{k(\bm{Z}^{(t)})w(\bm{Z}^{*})}= \dfrac{f_{\bm{\psi^{(t)}}}(\bm{Z}^{*}|\bm{Y})g_{\bm{\alpha}^{(t)}}(\bm{Z}^{(t)})}{f_{\bm{\psi}^{(t)}}(\bm{Z}^{(t)}|\bm{Y})g_{\bm{\alpha}^{(t)}}(\bm{Z}^{*})}
 \varpropto \dfrac{f_{\bm{\psi^{(t)}}}(\bm{Y}|\bm{Z}^{*})g_{\bm{\alpha}^{(t)}}(\bm{Z}^{*})g_{\bm{\alpha}^{(t)}}(\bm{Z}^{(t)})}{f_{\bm{\psi^{(t)}}}(\bm{Y}|\bm{Z}^{(t)})g_{\bm{\alpha}^{(t)}}(\bm{Z}^{t})g_{\bm{\alpha}^{(t)}}(\bm{Z}^{*})}\notag
\end{eqnarray}
Which yields:
\begin{equation}
R(\bm{Z}^{(t)},\bm{Z}^{*})=\dfrac{f_{\bm{\psi^{(t)}}}(\bm{Y}|\bm{Z}^{*})}{f_{\bm{\psi^{(t)}}}(\bm{Y}|\bm{Z}^{(t)})}
\end{equation}
Finally, we find sample $Z^{(t+1)}$ using following equation:
\begin{equation}
        \bm{Z}^{(t+1)}= \begin{cases}
            \bm{Z}^{*} & \text{with probability min$\{R\left(\bm{Z}^{(t)}, \bm{Z}^{*}\right),1 \}$} \\
            \bm{Z}^{(t)} & \text{otherwise}.
            \end{cases}\label{next-sample2}
        \end{equation}\label{Hastings2}

In the $t^\text{th}$ iteration of the EM algorithm, we use MH algorithm to generate $10000$ independent sample points $Z^{(t,1)}, Z^{(t,2)},..., Z^{(t,10000)} $ from probability distribution $f_{\psi^{(t)}}(Z|Y)$, as showed above. Note that we remove the first 10000 samples generated by MH algorithm to make sure the MH algorithm converges, and keep samples with 500 gaps to make sure the selected samples are independent.

\normalsize
\textbf{M-Step}:

We maximize  $Q_{MC}(\bm{\psi},\bm{\psi}^{(t)})$ with respect to the parameters $\bm{\psi}=(\bm{\theta},\bm{\alpha})$. We can see that 
$\widehat{E}_{\bm{\psi}^{(t)}}\{\log {f(\bm{Y}|\bm{Z};\bm{\theta})|\bm{Y}}\}$ is a function of $\bm{\theta}$ only. We can optimize the parameters using constrOptim in R. $\widehat{E}_{\bm{\psi^{(t)}}}\{\log {g(\bm{Z};\bm{\alpha})|\bm{Y}}\}$ is a function of $\bm{\alpha}$ only. We can use dirichlet.mle function in R to optimize them. We stop the MCEM algorithm when sum of the absolute values of differences between parameters in consecutive iterations is smaller than $0.01$.

\section{Simulation}\label{simulation}
\begin{table}
\centering
\caption{Eight simulation settings of relative risks and eight scenarios with three factors. }
\noindent\adjustbox{max width=\textwidth}{
\begin{tabular}{l l}

\begin{tabular}{c c c c c c c}
\hline
\hline 
\multicolumn{4}{c}{} A. Disease Model& \\
\cline{1-6}
Models & $R_1$ & $R_2$ & $R_{im}$ & $S_1$ & $S_2$ \\ [0.9ex] 
\hline 
1 & 1 & 1 & 1  & 1 & 1 \\ 
2 & 2 & 3 & 1  & 1 & 1  \\
3 & 1 & 3 & 1  & 1 & 1   \\
4 & 1 & 3 & 1  & 2 &  2  \\
5 & 1 & 3 & 3  & 1 & 1   \\  
6 & 3 & $3$ & 1/3& 1& 1   \\
7 & 1 & 3 & 3  & 2 &2    \\
8 & 3 & $3$ & 1/3& 2&2  \\[0.5ex]
\hline 
\end{tabular}
&

\begin{tabular}{c c  c c}
\hline
\hline 
\multicolumn{3}{c}{}B. Factor Values&  \\
\cline{1-4}
Scenarios &  MAF & PREV & HWE \\ [0.9ex] 
\hline 
1 & 0.1  & 0.05 & 0  \\ 
2 & 0.1  & 0.05 & 1  \\
3 & 0.1  & 0.15 & 0  \\
4 &  0.1  & 0.15 & 1  \\
5 &0.3 & 0.05 & 0  \\  
6 & 0.3 & 0.05 & 1   \\
7 & 0.3 & 0.15 & 0   \\
8 & 0.3 &  0.15& 1  \\[0.5ex]
\hline 
\end{tabular}
\end{tabular}}
\label{table:nonlin} 
\end{table}

We simulate eight disease models and scenarios according to table \ref{table:nonlin} to examine and compare the power of our MCEM method with  LIME. In table \ref{table:nonlin}, the first disease model corresponds to no genetic effect, models 2 and 3 correspond to the genetic effects that are not caused by either imprinting nor maternal effect, model 4 corresponds to only maternal effect, models 5 and 6 correspond to imprinting, and models 7 and 8 correspond both maternal effect and imprinting. Following table \ref{table:nonlin}, under each model, we run eight different scenarios spanned by three factor values each with two levels: minor allele frequency (MAF) at $0.1$ and $0.3$, disease prevalence $P(D=1)$ (PREV) at $0.05$(rare), $0.15$ (common), and whether Hardy-Weinberg equilibrium (HWE) holds at $1=$Yes and $0=$No. Denote $p$ as minor allele frequency. When Hardy-Weinberg holds, the probabilities of genotypes containing $0$, $1$, and $2$ minor alleles are $(1-p)^2$, $2p(1-p)$, and $(1-p)^2$, respectively. As a result of population HWE, allelic exchangeability (AE) and mating symmetry (MS) will be implied. When HWE is violated, the probabilities of genotypes with $0$, $1$, and $2$ minor alleles are $(1-p)^2(1-\zeta)+(1-p)(1-\zeta)$, $2p(1-p)(1-\zeta)$, and $p^2(1-\zeta)+p\zeta$, where $\zeta$ is the inbreeding coefficient parameter \citep{lynch1998genetics}. Note that $\zeta$ is set to be $0.1$ for males and $0.3$ for females. When HWE does not hold, both AE and MS are not satisfied.

In the simulation, parents' genotypes are generated according to genotype probabilities. Children's genotypes are sampled based on the transmission probability. Children's disease statuses are generated based on calculated disease penetrance rate in (\ref{parameter1}). We repeat the process until obtaining $150$ case-parent and $150$ control-parent triads. $15$ types of child-parent triads among the case and control families are counted. Each simulation setting is repeatedly simulated for $500$ times.

\subsection{Hypothesis Testing}\label{test}

We test association, imprinting, and maternal effects based on log likelihood ratio tests. For association test,
\[
H_0: R_1=R_2=R_{im}=S_1=S_2=1 \quad \text{vs.} \quad H_1: \text{at least one of these parameters is not 1.}
\]

For imprinting effect,
\[
H_0: R_{im}=1 \quad \text{vs.} \quad H_1: R_{im} \neq 1
\]

For testing the role of maternal effect:
\[
H_0: S_1=S_2=1 \quad \text{vs.} \quad H_1: \text{$S_1\ne 1$ or $S_2\ne 1$.}
\]

We apply MCEM algorithm to both null model and alternative model. When the algorithm converges, we denote the monte carlo estimation $\widehat{E}_{\psi^{(t)}}\left\{\log {f(Y|Z)} \right\}$ in the last iteration as $\widehat{E}_{\psi_{0}^{(t)}}\left\{\log {f(Y|Z)}  \right\}$ and $\widehat{E}_{\psi_1^{(t)}}\left\{\log {f(Y|Z)} \right\}$  for null model and alternative model, respectively. To test each hypothesis we calculate log likelihood ratio as follow:
\begin{equation}\label{log-ratio}
LR = -2 \log \left[ \dfrac{\widehat{E}_{\psi_{0}^{(t)}}\left\{\log {f(Y|Z)}  \right\}}{\widehat{E}_{\psi_{1}^{(t)}}\left\{\log {f(Y|Z)} \right\}}  \right].
\end{equation}
Reject null hypothesis at $\alpha$ significance level if test statistic is greater than Chi-square distribution $\chi^2_{d,\alpha}$, where degree of freedom $d$ is the difference of the number of free parameters between alternative model and null model. For association test, when there are no additional siblings, as the first two terms in $\log f(\bm{Y}|\bm{Z};\theta)$ shown in (\ref{derive11}) under null hypothesis are free of $\bm{\theta}$, $d$ is 6; when there are additional siblings, as $\log f(\bm{Y}|\bm{Z};\theta)$ under null hypothesis depends on $\bm{\delta}$, $d$ is 5. For imprinting test, $R_{im}=1$ under null hypothesis, so $d$ is 1. For maternal effect, $S_1=S_2=1$ under null hypothesis, so $d$ is 2.

\subsection{Empirical type I error rate and power}\label{typeI}
\begin{figure}[H]
\centering
\subfloat[Subfigure 5 list of figures text][]{
\includegraphics[width=0.4\textwidth]{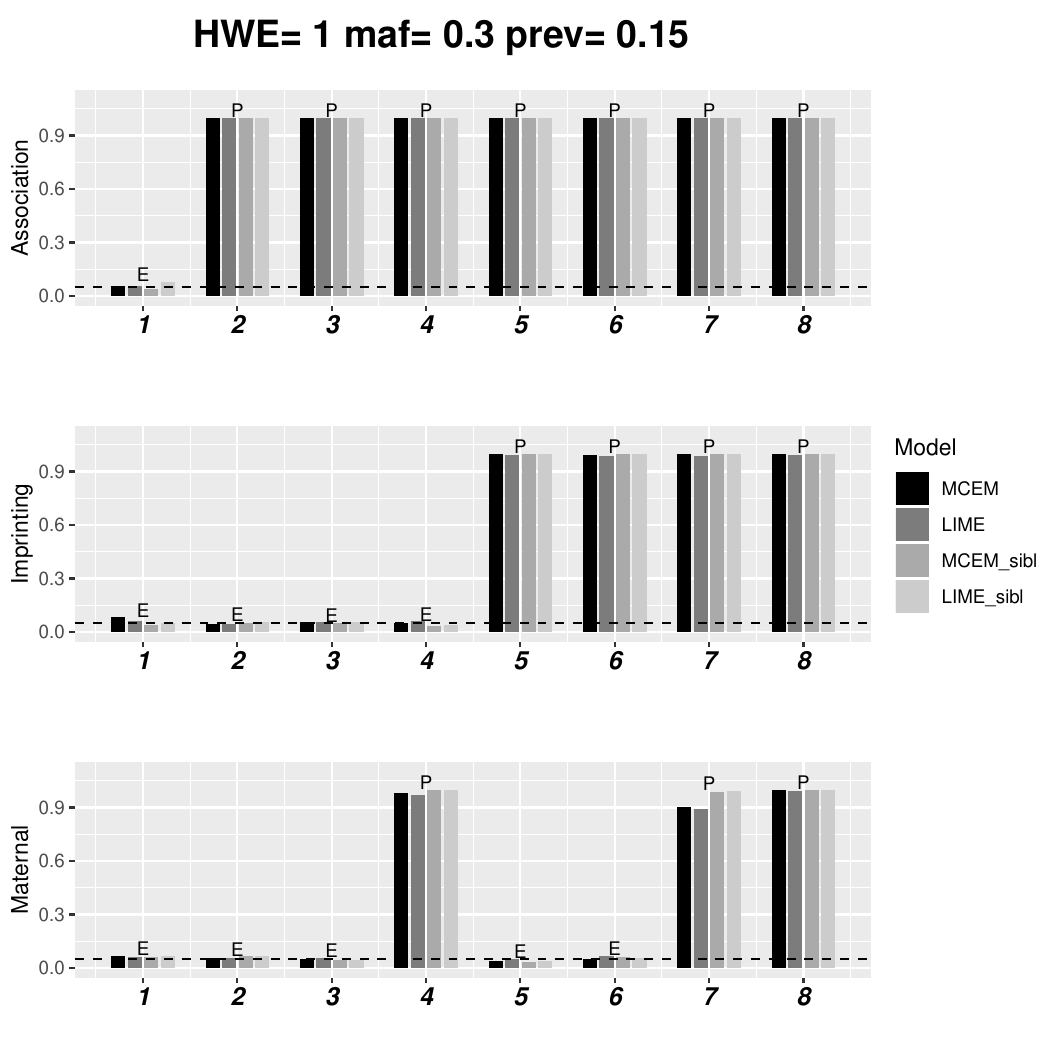}
\label{fig:subfig1}}
\qquad
\subfloat[Subfigure 6 list of figures text][]{
\includegraphics[width=0.4\textwidth]{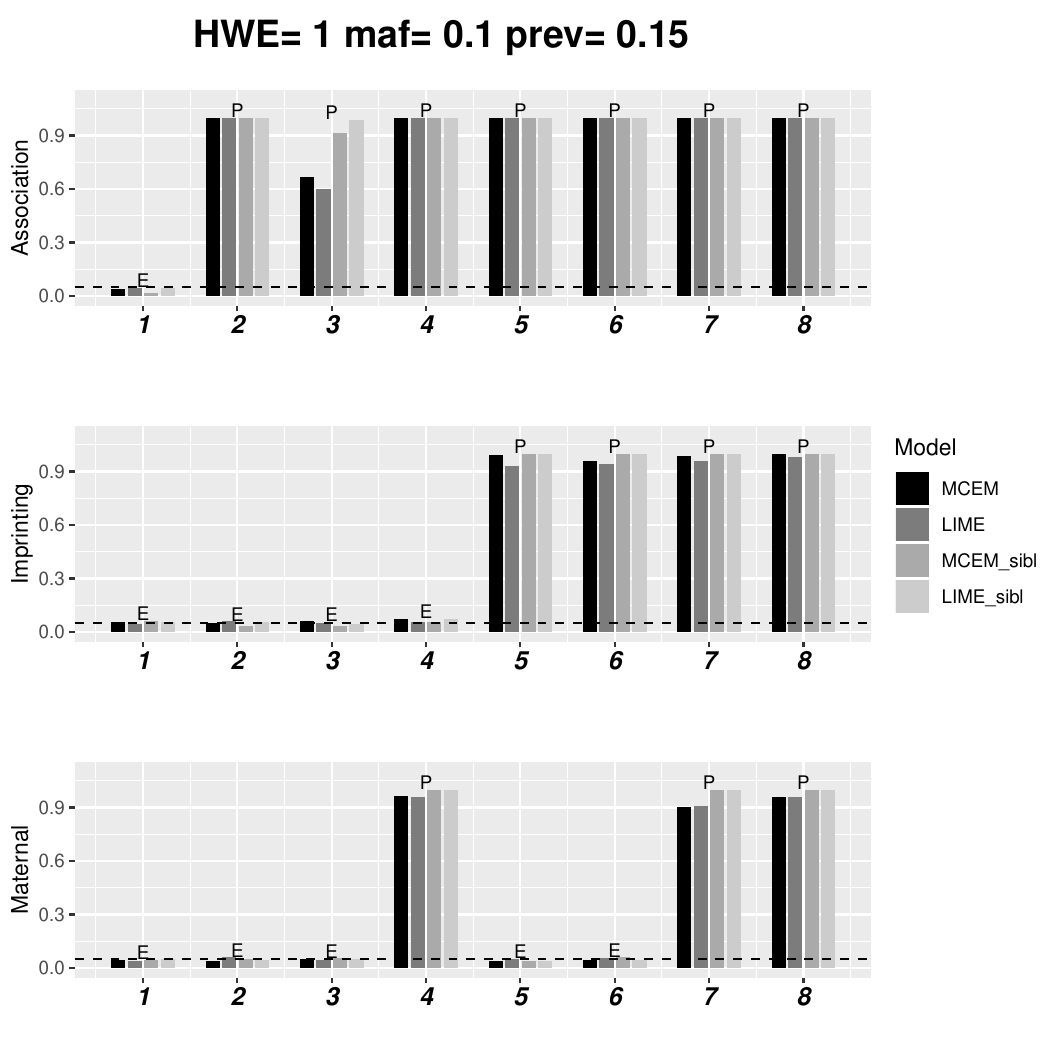}
\label{fig:subfig2}}
\qquad
\subfloat[Subfigure 7 list of figures text][]{
\includegraphics[width=0.4\textwidth]{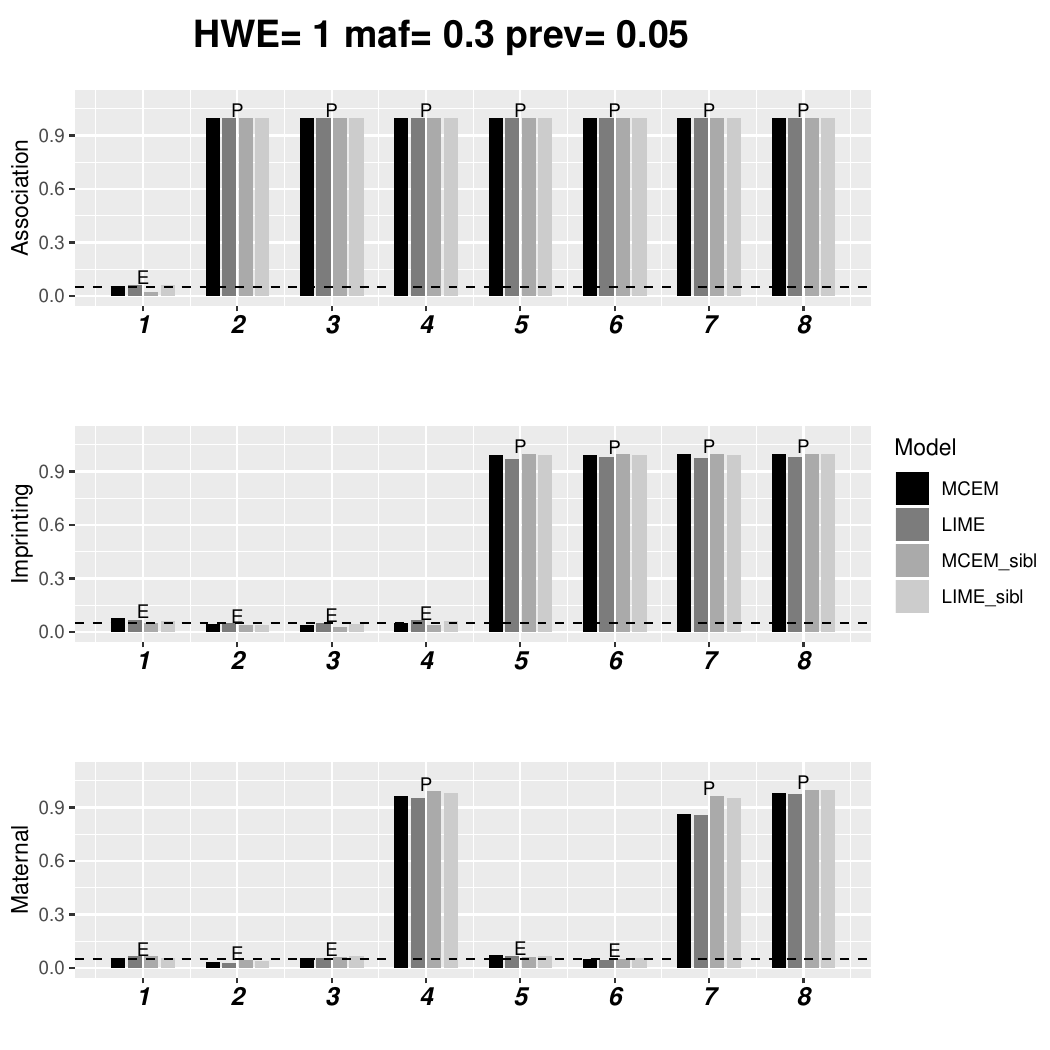}
\label{fig:subfig3}}
\qquad
\subfloat[Subfigure 8 list of figures text][]{
\includegraphics[width=0.4\textwidth]{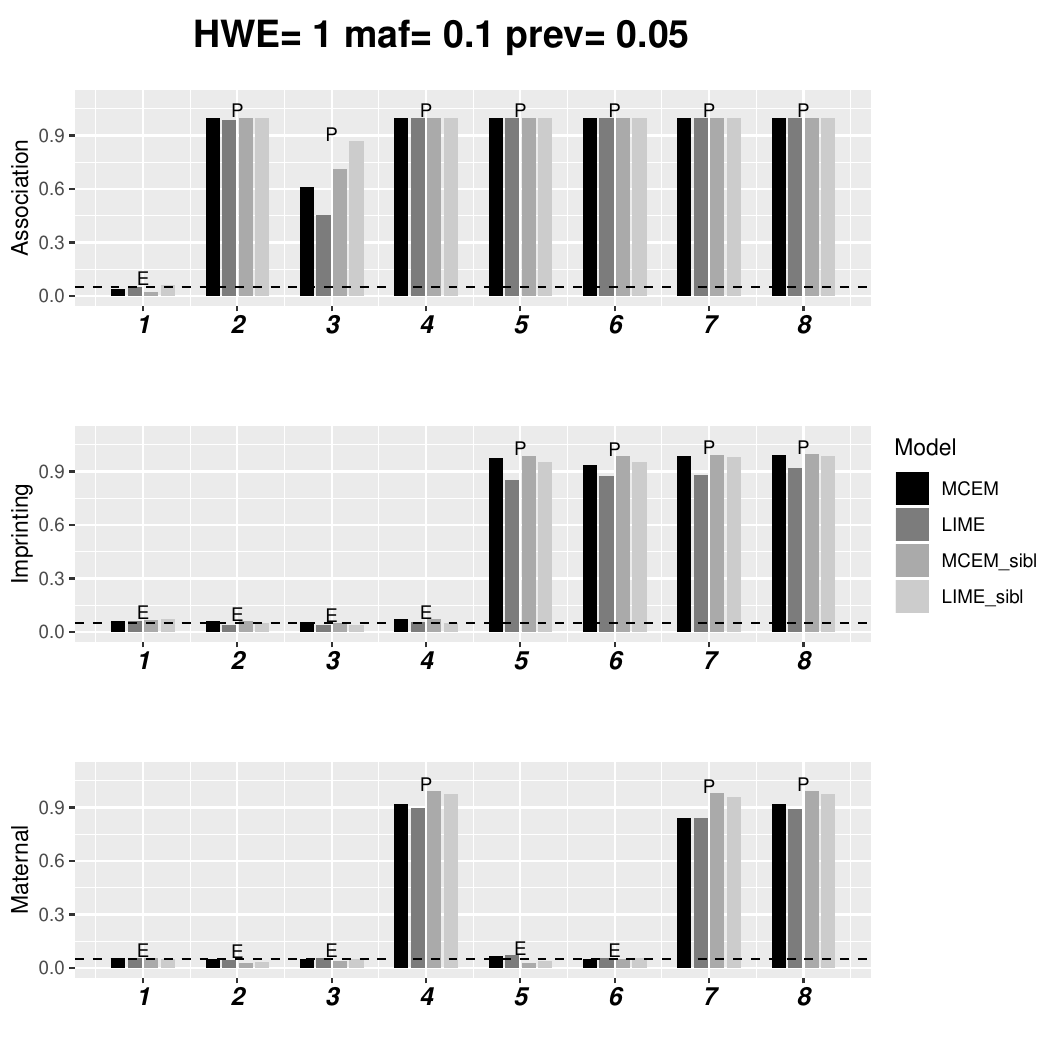}
\label{fig:subfig4}}
\caption{Barplots for type 1 error under HWE=1.}
\label{fig:globfig5}
\end{figure}

We compared the type I error rates
and power of MCEM and LIME in families with and without additional siblings under each setting. 
The results in Figures (\ref{fig:globfig4}) and (\ref{fig:globfig5}) show that MCEM method can control type I error well. As expected, the power of MCEM for data with additional siblings is higher than that without additional siblings. The power of MCEM for association and imprinting effects are generally higher than LIME. The power of MCEM for maternal effect is generally lower than LIME when there are no additional siblings, but on the other hand, when there are additional siblings, MCEM is generally more powerful in detecting maternal effect.

\subsection{Wild Estimates and Bias}\label{bias}
When a parameter estimate exceeding $20$ or falling below $1/20$, we consider it too far away from the true value, so we call it wild estimate. We computed the proportion of wild estimates for both MCEM and LIME methods. The results of wild estimate proportion for the data without additional siblings are represented in tables (\ref{table:nonlin11})...(\ref{table:nonlin18}).  The tables show that the proportions of wild estimates for LIME method are much higher than MCEM, especially when minor allele frequency is low. With additional siblings, MCEM also found much less significant proportion of wild estimates as well.

\begin{figure}[h]
\centering
\subfloat[Subfigure 1 list of figures text][]{
\includegraphics[width=0.4\textwidth]{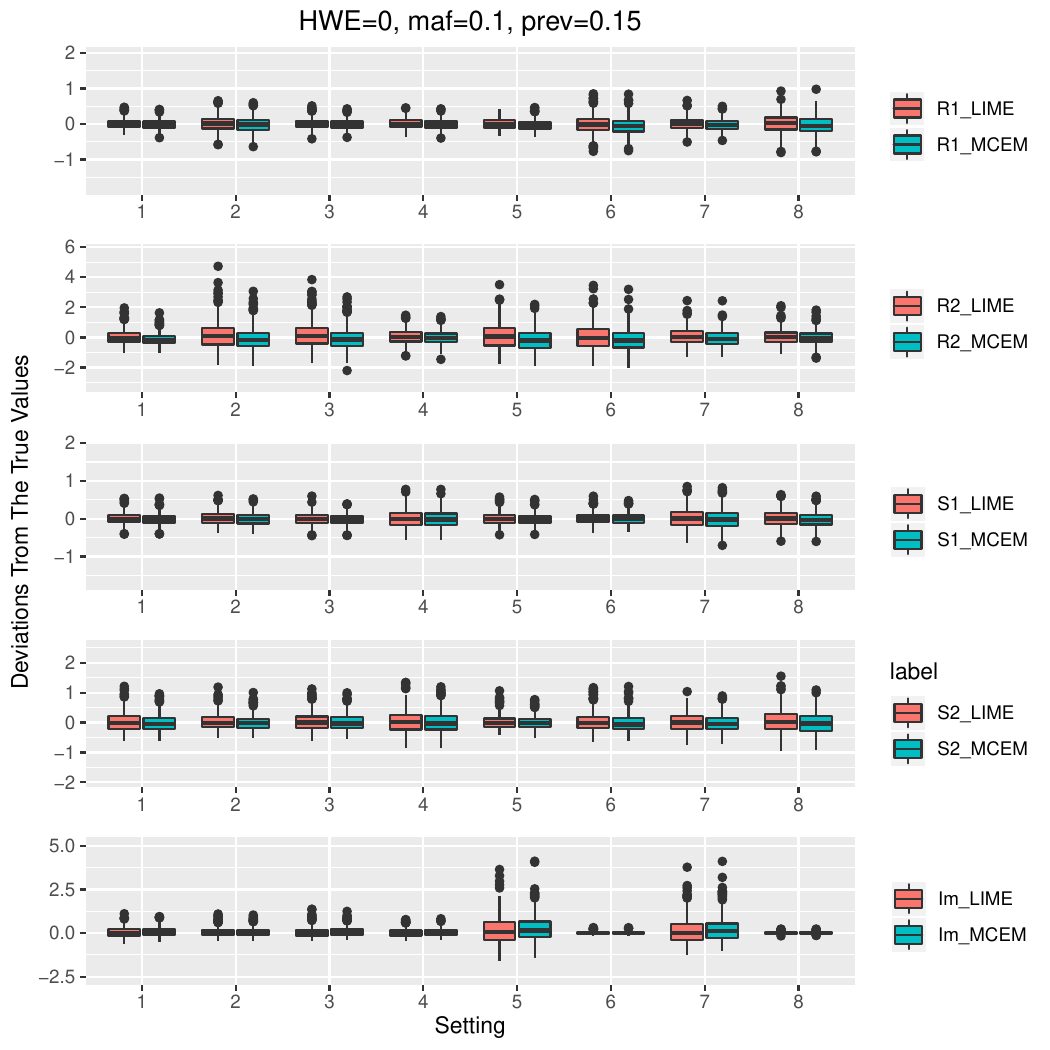}
\label{fig:subfig1}}
\qquad
\subfloat[Subfigure 2 list of figures text][]{
\includegraphics[width=0.4\textwidth]{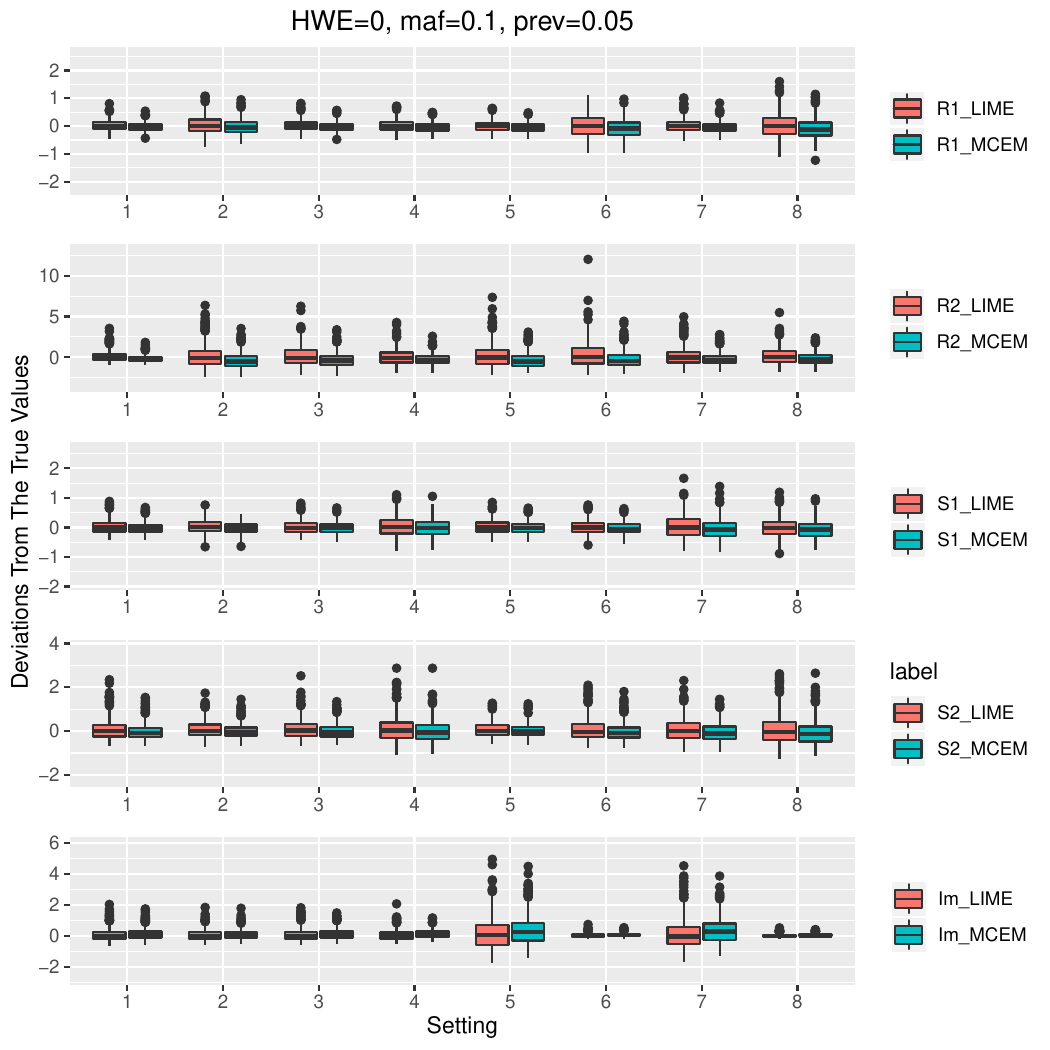}
\label{fig:subfig2}}
\qquad
\subfloat[Subfigure 3 list of figures text][]{
\includegraphics[width=0.4\textwidth]{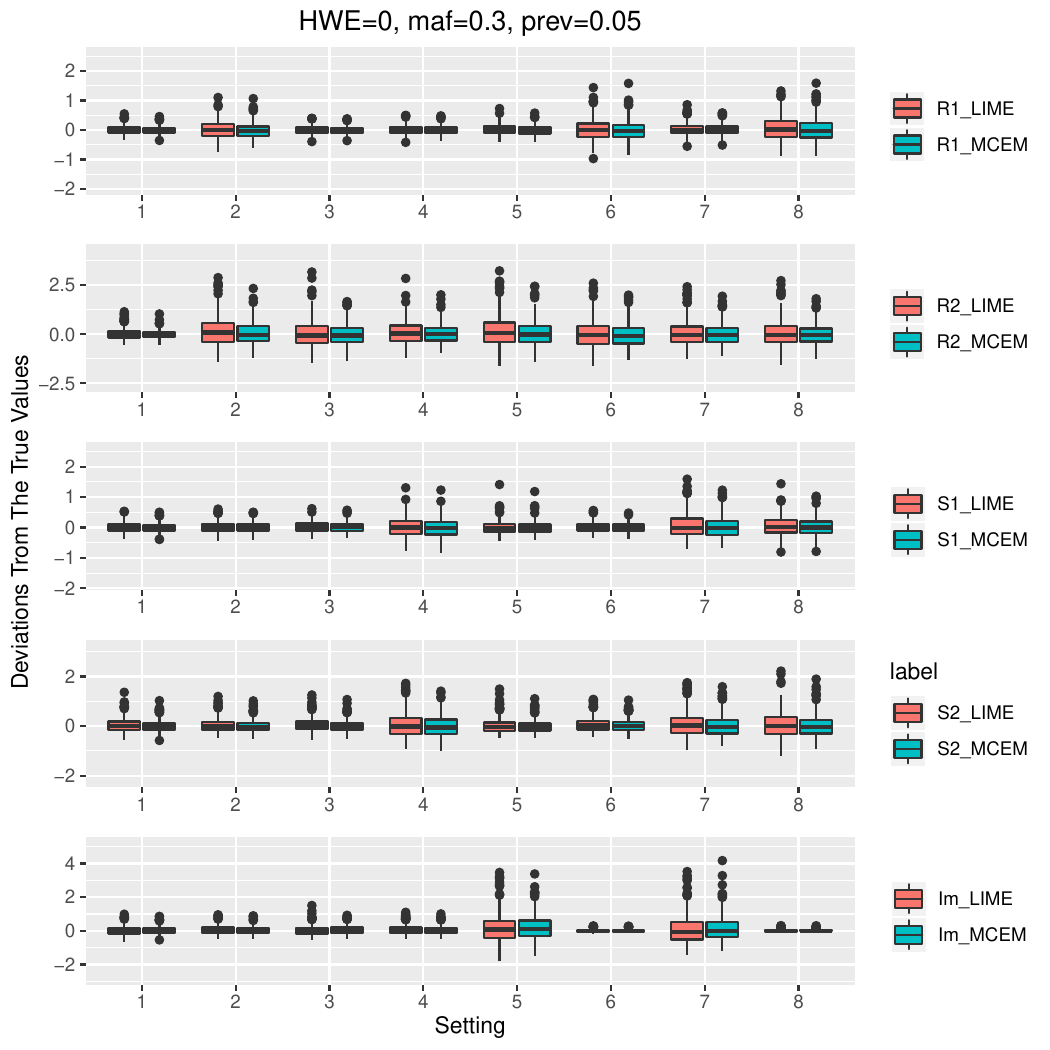}
\label{fig:subfig3}}
\qquad
\subfloat[Subfigure 4 list of figures text][]{
\includegraphics[width=0.4\textwidth]{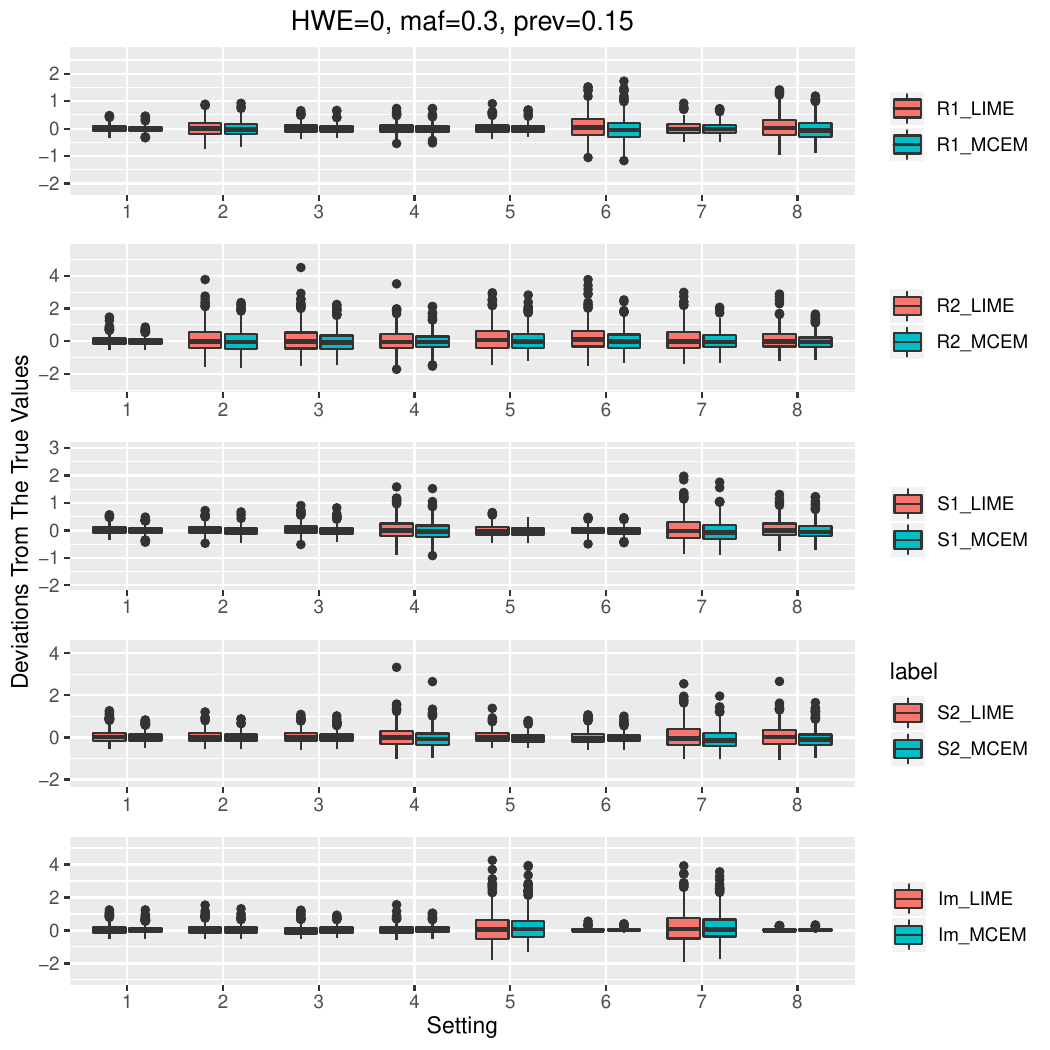}
\label{fig:subfig4}}
\caption{Box plots for biases of MCEM and LIME methods for the families without additional siblings where HWE=0.  }
\label{fig:globfig1}
\end{figure}

We calculated relative difference $\dfrac{(\widehat{\gamma}-\gamma)}{\gamma}$ for parameters $R_{1}$, $R_{2}$, $R_{im}$, $S_{1}$, and $S_{2}\}$. To make the results easier to read, we draw boxplots for relative difference for estimates excluding wild estimates (\ref{fig:globfig1})...(\ref{fig:globfig4}). From the figures, we can see that even wild estimates have been excluded, LIME estimates still have much larger standard error compared to MCEM estimates. On the other hand, the medians of LIME estimates are closer to the true values.  

\subsection{computation time}
MCEM algorithm takes much longer computational time than LIME method. For our computational work, we have access to the clusters in Texas Tech High Performance Computing Centers and to finish all the 500 replicates for one setting of  MCEM, it took around 3 hours, but for LIME it took less than a minute. In addition, we also found sometimes that MCEM algorithm can be sensitive to initial values for both $\bm{\alpha}$ and $\bm{\theta}$ and the similar issue has also been seen in LIME method. Further study is needed about the strategy to select the initial values.

\section{Conclusion}

 In this study, we proposed a Monte Carlo Expectation Maximization algorithm to detect imprinting effect and maternal effect for case-control family data. An extensive simulation has been used to demonstrate the efficiency of our methods. In particular, simulation results show that the MCEM algorithm can generally detect both epigenetic effects with higher power and smaller standard error compared to LIME method. However, LIME method can run much faster, and give parameter estimates with smaller bias.

\clearpage

\section*{Supplementary Materials}
\subsection{Tables and Figures}

\begin{table}
\centering
\caption{Wild estimation proportion by LIME versus MCEM for scenario HWE$=0$, maf$=0.1$, prev$=0.05$. }

\hspace{35mm} HWE$=0$, maf$=0.1$, prev$=0.05$  \\
\noindent\adjustbox{max width=\textwidth}{
\begin{tabular}{l l}
\begin{tabular}{c c c c c c c}

\cline{1-6}
\multicolumn{3}{c}{} &LIME  \\
\cline{2-6}
Settings & $R_1$ & $R_2$ & $R_{im}$ & $S_1$ & $S_2$ \\ [0.9ex] 
\hline 
1 & 0     & 0.326 & 0     & 0.04  & 0.008 \\
2 & 0     & 0.232 & 0.006 & 0.02  & 0.008 \\
3 & 0     & 0.158 & 0     & 0.04  & 0.008 \\
4 & 0     & 0.03  & 0     & 0.022 & 0.002 \\
5 & 0.002 & 0.246 & 0.01  & 0.02  & 0.064 \\
6 & 0     & 0.222 & 0.002 & 0.05  & 0.032 \\
7 & 0.004 & 0.118 & 0.002 & 0.01  & 0.026 \\
8 & 0     & 0.09  & 0     & 0.022 & 0.016\\[0.5ex]
\hline 
\end{tabular}
&

\begin{tabular}{ c c c c c  }

\cline{1-5}
\multicolumn{2}{c}{} &MCEM \\
\cline{1-5}
 $R_1$ & $R_2$ & $R_{im}$ & $S_1$ & $S_2$ \\ [0.9ex] 
\hline 
 0     & 0     & 0     & 0     & 0     \\
 0     & 0.004 & 0     & 0     & 0     \\
 0     & 0.004 & 0     & 0.006 & 0     \\
 0     & 0     & 0     & 0     & 0     \\
 0.002 & 0.018 & 0.008 & 0.008 & 0.022 \\
 0     & 0.006 & 0     & 0.002 & 0.002 \\
 0     & 0.002 & 0     & 0     & 0.008 \\
 0     & 0     & 0     & 0     & 0    \\[0.5ex]
\hline 
\end{tabular}
\end{tabular}}
\label{table:nonlin11} 
\end{table}

\begin{table}
\centering
\caption{Wild estimation proportion by LIME versus MCEM for scenario HWE$=0$, maf$=0.3$, prev$=0.05$. }

\hspace{35mm} HWE$=0$, maf$=0.05$, prev$=0.3$  \\
\noindent\adjustbox{max width=\textwidth}{
\begin{tabular}{l l}
\begin{tabular}{c c c c c c c}

\cline{1-6}
\multicolumn{3}{c}{} &LIME  \\
\cline{2-6}
Settings & $R_1$ & $R_2$ & $R_{im}$ & $S_1$ & $S_2$ \\ [0.9ex] 
\hline 
1 & 0     & 0     & 0 & 0 & 0     \\
2 & 0     & 0     & 0 & 0 & 0     \\
3 & 0     & 0.004 & 0 & 0 & 0     \\
4 & 0     & 0     & 0 & 0 & 0.002 \\
5 & 0     & 0.008 & 0 & 0 & 0.008 \\
6 & 0     & 0.012 & 0 & 0 & 0.008 \\
7 & 0.002 & 0.008 & 0 & 0 & 0.014 \\
8 & 0     & 0     & 0 & 0 & 0.004\\[0.5ex]
\hline 
\end{tabular}
&

\begin{tabular}{ c c c c c  }

\cline{1-5}
\multicolumn{2}{c}{} &MCEM \\
\cline{1-5}
 $R_1$ & $R_2$ & $R_{im}$ & $S_1$ & $S_2$ \\ [0.9ex] 
\hline 
0     & 0 & 0 & 0 & 0     \\
0     & 0 & 0 & 0 & 0     \\
0     & 0 & 0 & 0 & 0     \\
0     & 0 & 0 & 0 & 0     \\
0     & 0 & 0 & 0 & 0     \\
0     & 0 & 0 & 0 & 0     \\
0.002 & 0 & 0 & 0 & 0.004 \\
0     & 0 & 0 & 0 & 0     \\[0.5ex]
\hline 
\end{tabular}
\end{tabular}}
\label{table:nonlin12} 
\end{table}


\begin{table}
\centering
\caption{Extreme values estimated by LIME versus MCEM for scenario HWE$=0$, maf$=0.1$, prev$=0.15$. }
\hspace{35mm} HWE$=0$, maf$=0.15$, prev$=0.1$  \\
\noindent\adjustbox{max width=\textwidth}{
\begin{tabular}{l l}
\begin{tabular}{c c c c c c c}

\cline{1-6}
\multicolumn{3}{c}{} &LIME  \\
\cline{2-6}
Settings & $R_1$ & $R_2$ & $R_{im}$ & $S_1$ & $S_2$ \\ [0.9ex] 
\hline 
1 & 0     & 0.262 & 0.002 & 0.03  & 0.012 \\
2 & 0     & 0.078 & 0.004 & 0.004 & 0.006 \\
3 & 0     & 0.04  & 0.002 & 0.026 & 0.002 \\
4 & 0.002 & 0.002 & 0     & 0.002 & 0.002 \\
5 & 0     & 0.098 & 0.008 & 0.01  & 0.042 \\
6 & 0     & 0.14  & 0.008 & 0.026 & 0.02  \\
7 & 0.006 & 0.022 & 0     & 0     & 0.028 \\
8 & 0     & 0.004 & 0     & 0     & 0  \\  [0.5ex]
\hline 
\end{tabular}
&

\begin{tabular}{ c c c c c  }

\cline{1-5}
\multicolumn{2}{c}{} &MCEM \\
\cline{1-5}
 $R_1$ & $R_2$ & $R_{im}$ & $S_1$ & $S_2$ \\ [0.9ex] 
\hline 
0     & 0.006 & 0.002 & 0.006 & 0.002 \\
0     & 0.002 & 0     & 0     & 0     \\
0     & 0.002 & 0     & 0.008 & 0.002 \\
0.002 & 0.002 & 0     & 0     & 0.002 \\
0     & 0.01  & 0.002 & 0.004 & 0.02  \\
0     & 0.004 & 0     & 0.004 & 0.006 \\
0     & 0.002 & 0     & 0     & 0.014 \\
0     & 0.002 & 0     & 0     & 0.002\\[0.5ex]
\hline 
\end{tabular}
\end{tabular}}
\label{table:nonlin13} 
\end{table}


\begin{table}
\centering
\caption{Extreme values estimated by LIME versus MCEM for scenario HWE$=0$, maf$=0.15$, prev$=0.3$. }
\hspace{35mm} HWE$=0$, maf$=0.3$, prev$=0.15$  \\
\noindent\adjustbox{max width=\textwidth}{
\begin{tabular}{l l}
\begin{tabular}{c c c c c c c}

\cline{1-6}
\multicolumn{3}{c}{} &LIME  \\
\cline{2-6}
Settings & $R_1$ & $R_2$ & $R_{im}$ & $S_1$ & $S_2$ \\ [0.9ex] 
\hline 
1 & 0 & 0     & 0 & 0     & 0     \\
2 & 0 & 0.006 & 0 & 0.002 & 0     \\
3 & 0 & 0     & 0 & 0     & 0     \\
4 & 0 & 0     & 0 & 0     & 0     \\
5 & 0 & 0.01  & 0 & 0     & 0.022 \\
6 & 0 & 0.002 & 0 & 0     & 0.002 \\
7 & 0 & 0.008 & 0 & 0     & 0.018 \\
8 & 0 & 0     & 0 & 0     & 0     \\  [0.5ex]
\hline 
\end{tabular}
&

\begin{tabular}{ c c c c c  }

\cline{1-5}
\multicolumn{2}{c}{} &MCEM \\
\cline{1-5}
 $R_1$ & $R_2$ & $R_{im}$ & $S_1$ & $S_2$ \\ [0.9ex] 
\hline 
0 & 0     & 0 & 0 & 0     \\
0 & 0     & 0 & 0 & 0     \\
0 & 0     & 0 & 0 & 0     \\
0 & 0     & 0 & 0 & 0     \\
0 & 0.002 & 0 & 0 & 0.01  \\
0 & 0     & 0 & 0 & 0     \\
0 & 0     & 0 & 0 & 0.008 \\
0 & 0     & 0 & 0 & 0   \\[0.5ex]
\hline 
\end{tabular}
\end{tabular}}
\label{table:nonlin14} 
\end{table}


\begin{table}
\centering
\caption{Extreme values estimated by LIME versus MCEM for scenario HWE$=1$, maf$=0.1$, prev$=0.05$. }

\hspace{35mm} HWE$=1$, maf$=0.1$, prev$=0.05$  \\
\noindent\adjustbox{max width=\textwidth}{
\begin{tabular}{l l}
\begin{tabular}{c c c c c c c}

\cline{1-6}
\multicolumn{3}{c}{} &LIME  \\
\cline{2-6}
Settings & $R_1$ & $R_2$ & $R_{im}$ & $S_1$ & $S_2$ \\ [0.9ex] 
\hline 
1 & 0     & 0.34  & 0     & 0.348 & 0.004 \\
2 & 0     & 0.23  & 0     & 0.22  & 0.004 \\
3 & 0     & 0.174 & 0     & 0.234 & 0.004 \\
4 & 0     & 0.036 & 0     & 0.16  & 0     \\
5 & 0     & 0.238 & 0.002 & 0.104 & 0.022 \\
6 & 0.002 & 0.272 & 0     & 0.31  & 0.02  \\
7 & 0.004 & 0.078 & 0.002 & 0.058 & 0.032 \\
8 & 0     & 0.064 & 0     & 0.188 & 0.008 \\  [0.5ex]
\hline 
\end{tabular}
&

\begin{tabular}{ c c c c c  }

\cline{1-5}
\multicolumn{2}{c}{} &MCEM \\
\cline{1-5}
 $R_1$ & $R_2$ & $R_{im}$ & $S_1$ & $S_2$ \\ [0.9ex] 
\hline 
0 & 0     & 0 & 0 & 0     \\
0 & 0     & 0 & 0 & 0     \\
0 & 0     & 0 & 0 & 0     \\
0 & 0     & 0 & 0 & 0     \\
0 & 0.002 & 0 & 0 & 0.01  \\
0 & 0     & 0 & 0 & 0     \\
0 & 0     & 0 & 0 & 0.008 \\
0 & 0     & 0 & 0 & 0   \\[0.5ex]
\hline 
\end{tabular}
\end{tabular}}
\label{table:nonlin15} 
\end{table}


\begin{table}
\centering
\caption{Extreme values estimated by LIME versus MCEM for scenario HWE$=1$, maf$=0.3$, prev$=0.05$. }

\hspace{35mm} HWE$=1$, maf$=0.3$, prev$=0.05$  \\
\noindent\adjustbox{max width=\textwidth}{
\begin{tabular}{l l}
\begin{tabular}{c c c c c c c}

\cline{1-6}
\multicolumn{3}{c}{} &LIME  \\
\cline{2-6}
Settings & $R_1$ & $R_2$ & $R_{im}$ & $S_1$ & $S_2$ \\ [0.9ex] 
\hline 
1 & 0     & 0     & 0 & 0     & 0     \\
2 & 0     & 0     & 0 & 0.002 & 0     \\
3 & 0     & 0     & 0 & 0     & 0     \\
4 & 0     & 0.002 & 0 & 0     & 0     \\
5 & 0     & 0.004 & 0 & 0     & 0     \\
6 & 0     & 0.01  & 0 & 0     & 0.004 \\
7 & 0.002 & 0     & 0 & 0     & 0.012 \\
8 & 0     & 0     & 0 & 0.002 & 0.006 \\  [0.5ex]
\hline 
\end{tabular}
&

\begin{tabular}{ c c c c c  }

\cline{1-5}
\multicolumn{2}{c}{} &MCEM \\
\cline{1-5}
 $R_1$ & $R_2$ & $R_{im}$ & $S_1$ & $S_2$ \\ [0.9ex] 
\hline 
0 & 0 & 0 & 0 & 0     \\
0 & 0 & 0 & 0 & 0     \\
0 & 0 & 0 & 0 & 0     \\
0 & 0 & 0 & 0 & 0     \\
0 & 0 & 0 & 0 & 0     \\
0 & 0 & 0 & 0 & 0     \\
0 & 0 & 0 & 0 & 0.006 \\
0 & 0 & 0 & 0 & 0     \\[0.5ex]
\hline 
\end{tabular}
\end{tabular}}
\label{table:nonlin16} 
\end{table}


\begin{table}
\centering
\caption{Extreme values estimated by LIME versus MCEM for scenario HWE$=1$, maf$=0.1$, prev$=0.15$. }

\hspace{35mm} HWE$=1$, maf$=0.1$, prev$=0.15$  \\
\noindent\adjustbox{max width=\textwidth}{
\begin{tabular}{l l}
\begin{tabular}{c c c c c c c}

\cline{1-6}
\multicolumn{3}{c}{} &LIME  \\
\cline{2-6}
Settings & $R_1$ & $R_2$ & $R_{im}$ & $S_1$ & $S_2$ \\ [0.9ex] 
\hline 
1 & 0 & 0.258 & 0.002 & 0.254 & 0.004 \\
2 & 0 & 0.076 & 0     & 0.088 & 0     \\
3 & 0 & 0.026 & 0     & 0.212 & 0     \\
4 & 0 & 0.002 & 0     & 0.078 & 0     \\
5 & 0 & 0.042 & 0.002 & 0.04  & 0.022 \\
6 & 0 & 0.064 & 0     & 0.226 & 0.01  \\
7 & 0 & 0.01  & 0     & 0.014 & 0.01  \\
8 & 0 & 0.008 & 0     & 0.222 & 0   \\  [0.5ex]
\hline 
\end{tabular}
&

\begin{tabular}{ c c c c c  }

\cline{1-5}
\multicolumn{2}{c}{} &MCEM \\
\cline{1-5}
 $R_1$ & $R_2$ & $R_{im}$ & $S_1$ & $S_2$ \\ [0.9ex] 
\hline 
0 & 0.002 & 0.002 & 0.006 & 0     \\
0 & 0.002 & 0     & 0.006 & 0     \\
0 & 0     & 0     & 0.002 & 0     \\
0 & 0     & 0     & 0     & 0     \\
0 & 0.002 & 0.002 & 0.01  & 0.008 \\
0 & 0.002 & 0     & 0.002 & 0.002 \\
0 & 0     & 0     & 0     & 0.004 \\
0 & 0     & 0     & 0     & 0.002\\[0.5ex]
\hline 
\end{tabular}
\end{tabular}}
\label{table:nonlin17} 
\end{table}


\begin{table}
\centering
\caption{Extreme values estimated by LIME versus MCEM for scenario HWE$=1$, maf$=0.3$, prev$=0.15$. }

\hspace{35mm} HWE$=1$, maf$=0.3$, prev$=0.15$  \\
\noindent\adjustbox{max width=\textwidth}{
\begin{tabular}{l l}
\begin{tabular}{c c c c c c c}

\cline{1-6}
\multicolumn{3}{c}{} &LIME  \\
\cline{2-6}
Settings & $R_1$ & $R_2$ & $R_{im}$ & $S_1$ & $S_2$ \\ [0.9ex] 
\hline 
1 & 0 & 0 & 0 & 0     & 0     \\
2 & 0 & 0 & 0 & 0     & 0     \\
3 & 0 & 0 & 0 & 0     & 0     \\
4 & 0 & 0 & 0 & 0     & 0     \\
5 & 0 & 0 & 0 & 0     & 0.002 \\
6 & 0 & 0 & 0 & 0.002 & 0.002 \\
7 & 0 & 0 & 0 & 0     & 0.01  \\
8 & 0 & 0 & 0 & 0     & 0     \\  [0.5ex]
\hline 
\end{tabular}
&

\begin{tabular}{ c c c c c  }

\cline{1-5}
\multicolumn{2}{c}{} &MCEM \\
\cline{1-5}
 $R_1$ & $R_2$ & $R_{im}$ & $S_1$ & $S_2$ \\ [0.9ex] 
\hline 
0 & 0 & 0 & 0 & 0.002 \\
0 & 0 & 0 & 0 & 0     \\
0 & 0 & 0 & 0 & 0     \\
0 & 0 & 0 & 0 & 0     \\
0 & 0 & 0 & 0 & 0     \\
0 & 0 & 0 & 0 & 0     \\
0 & 0 & 0 & 0 & 0.002 \\
0 & 0 & 0 & 0 & 0       \\[0.5ex]
\hline 
\end{tabular}
\end{tabular}}
\label{table:nonlin18} 
\end{table}

\clearpage
\begin{figure}[h]
\centering
\subfloat[Subfigure 5 list of figures text][]{
\includegraphics[width=0.4\textwidth]{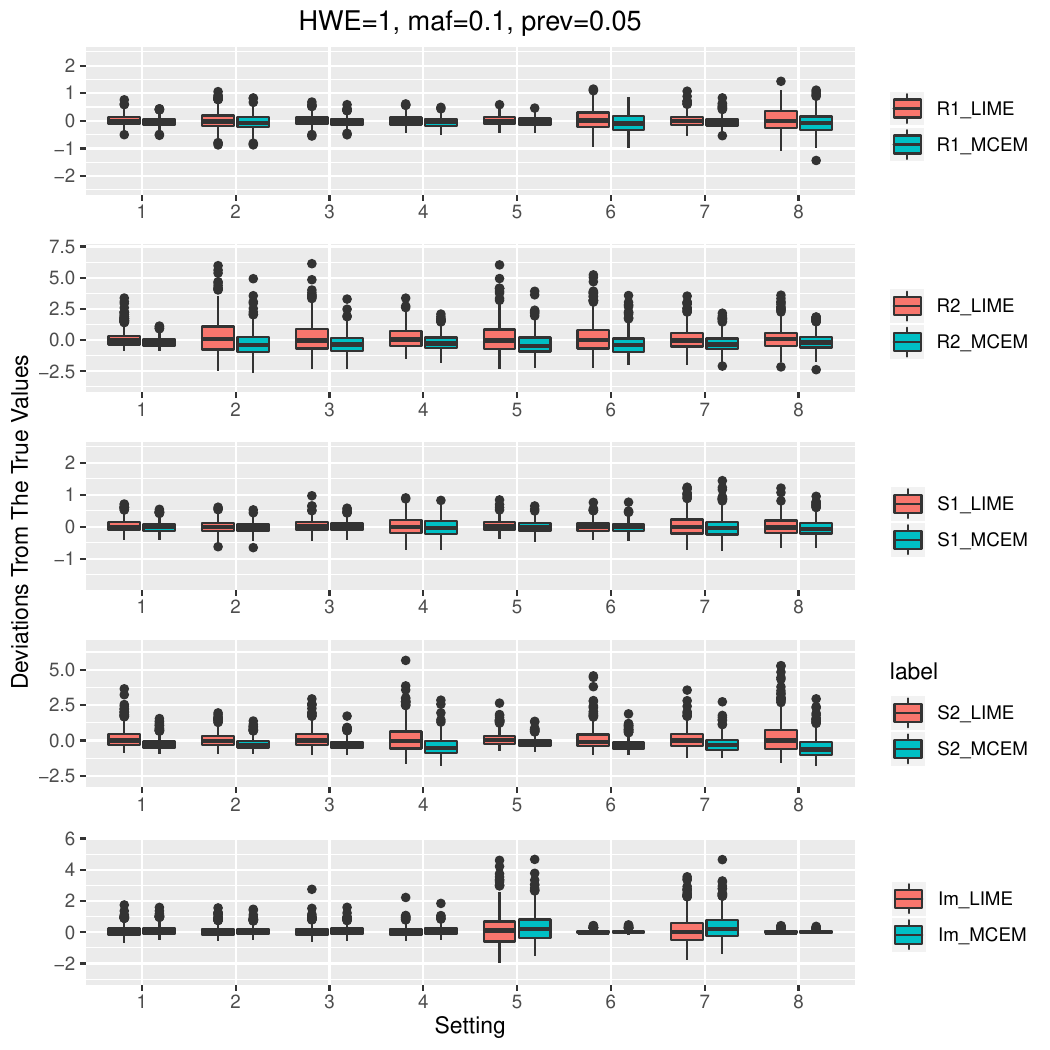}
\label{fig:subfig1}}
\qquad
\subfloat[Subfigure 6 list of figures text][]{
\includegraphics[width=0.4\textwidth]{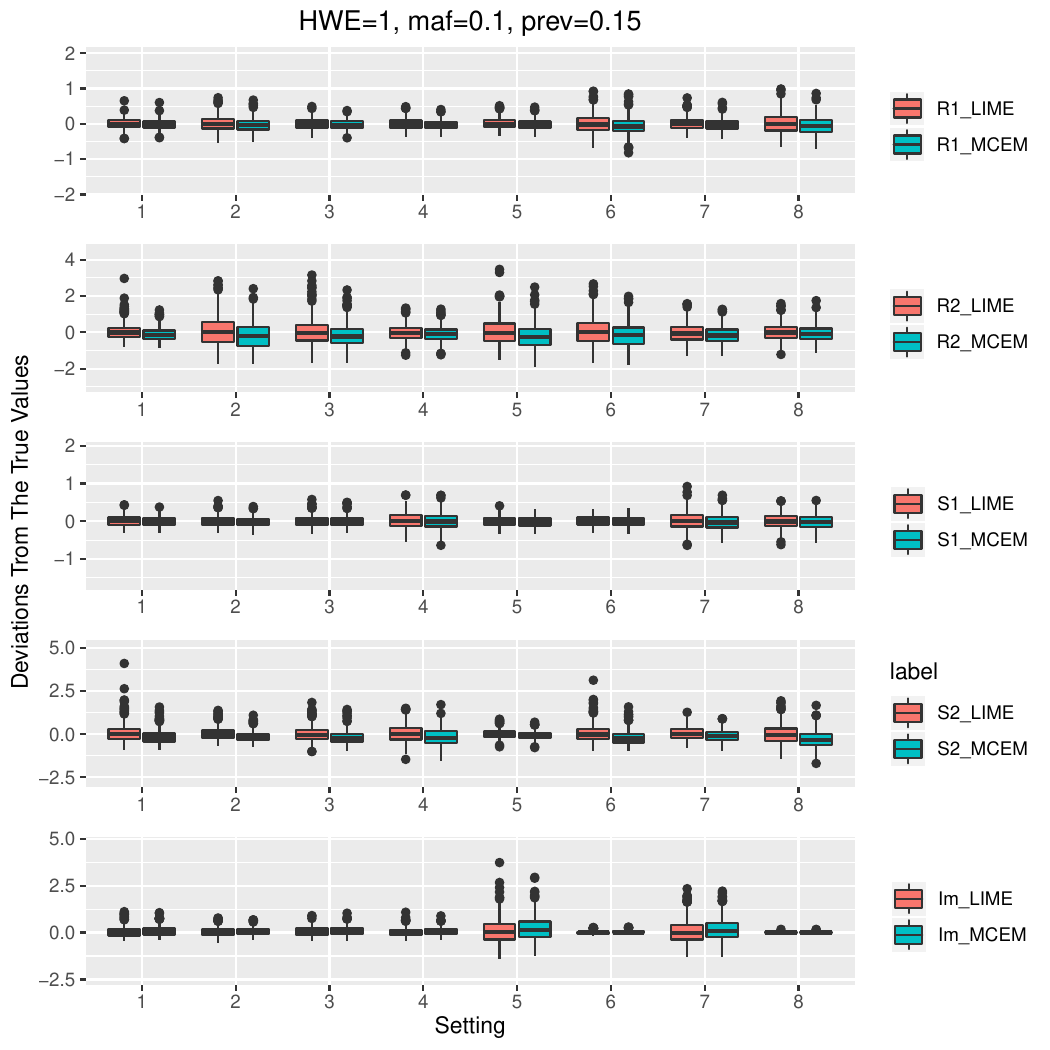}
\label{fig:subfig2}}
\qquad
\subfloat[Subfigure 7 list of figures text][]{
\includegraphics[width=0.4\textwidth]{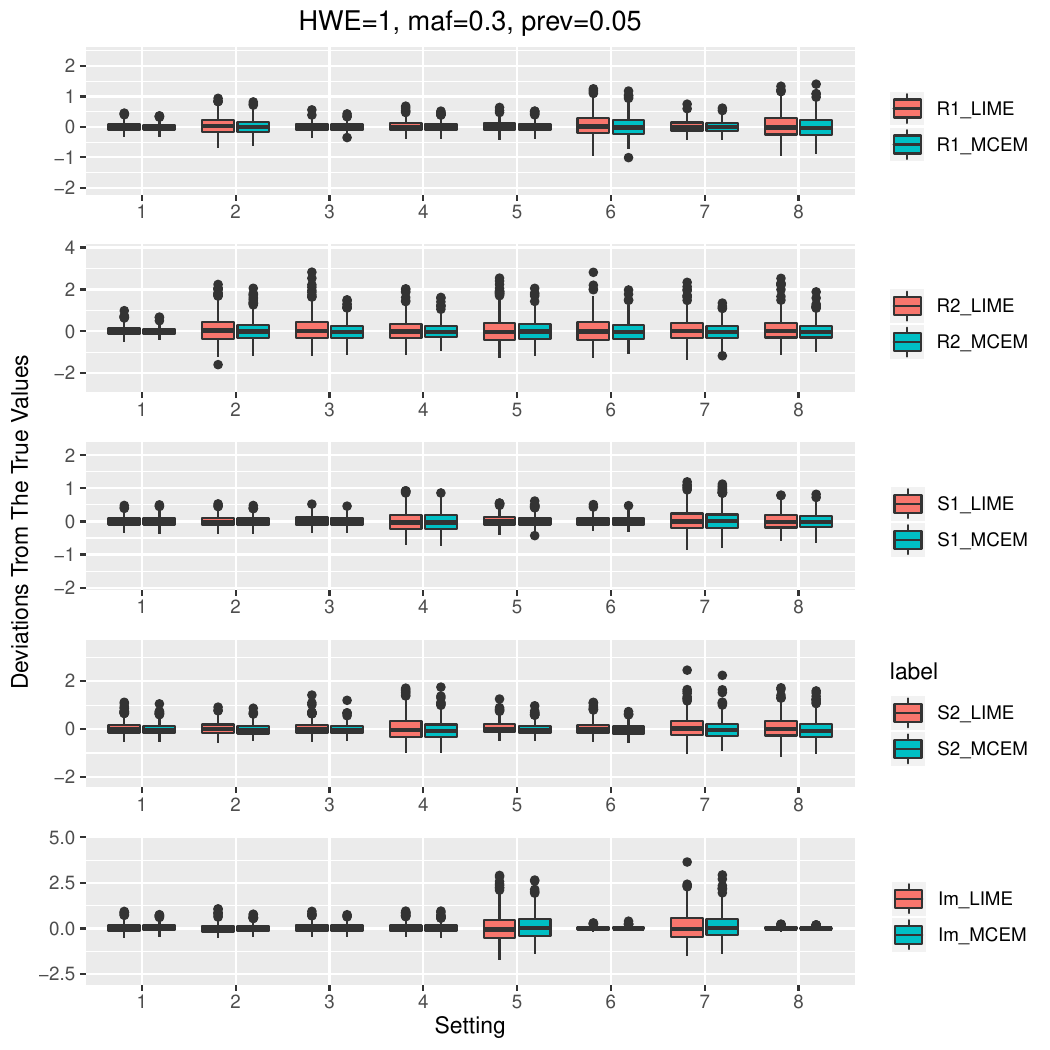}
\label{fig:subfig3}}
\qquad
\subfloat[Subfigure 8 list of figures text][]{
\includegraphics[width=0.4\textwidth]{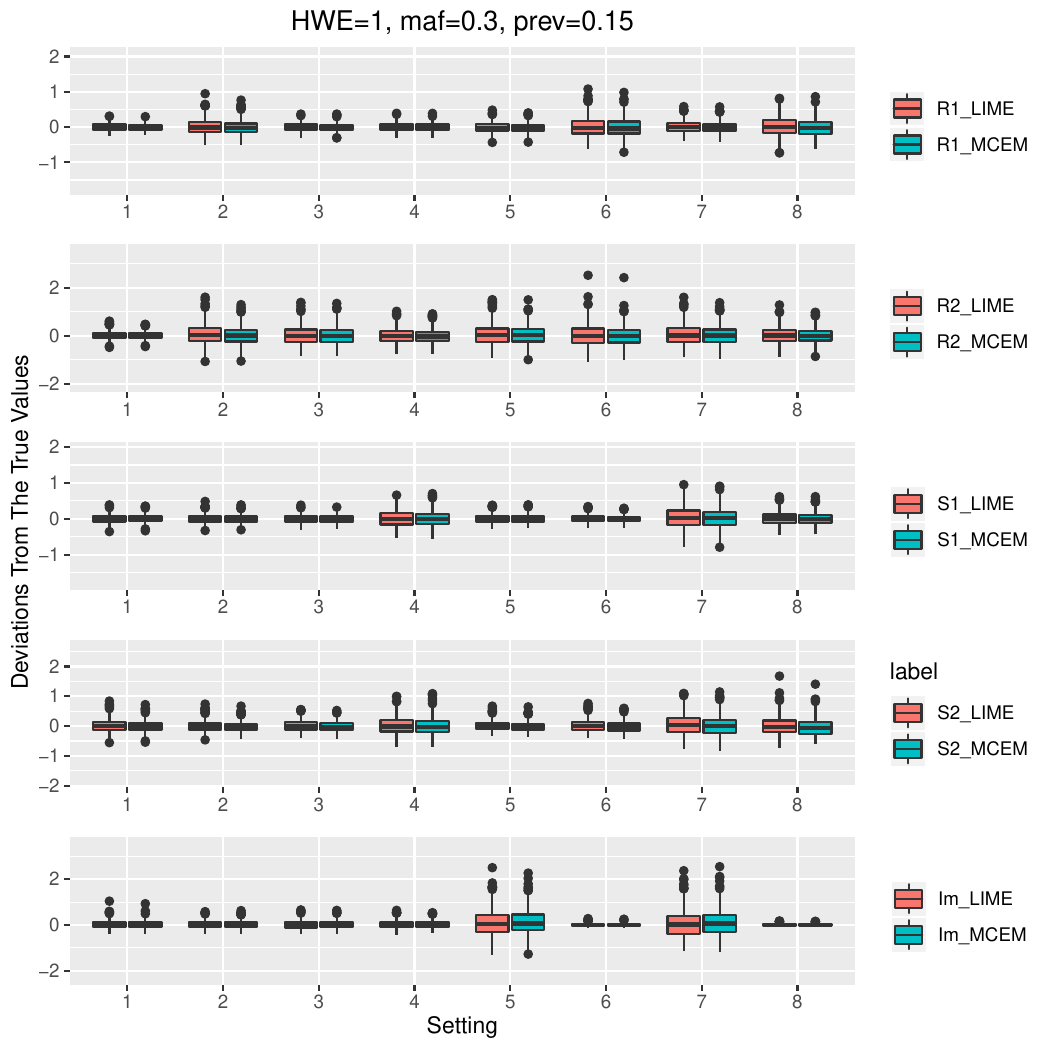}
\label{fig:subfig4}}
\caption{Box plots for biases of MCEM and LIME methods for the families without additional siblings where HWE=1.}
\label{fig:globfig2}
\end{figure}

\clearpage
\begin{figure}[h]
\centering
\subfloat[Subfigure 1 list of figures text][]{
\includegraphics[width=0.4\textwidth]{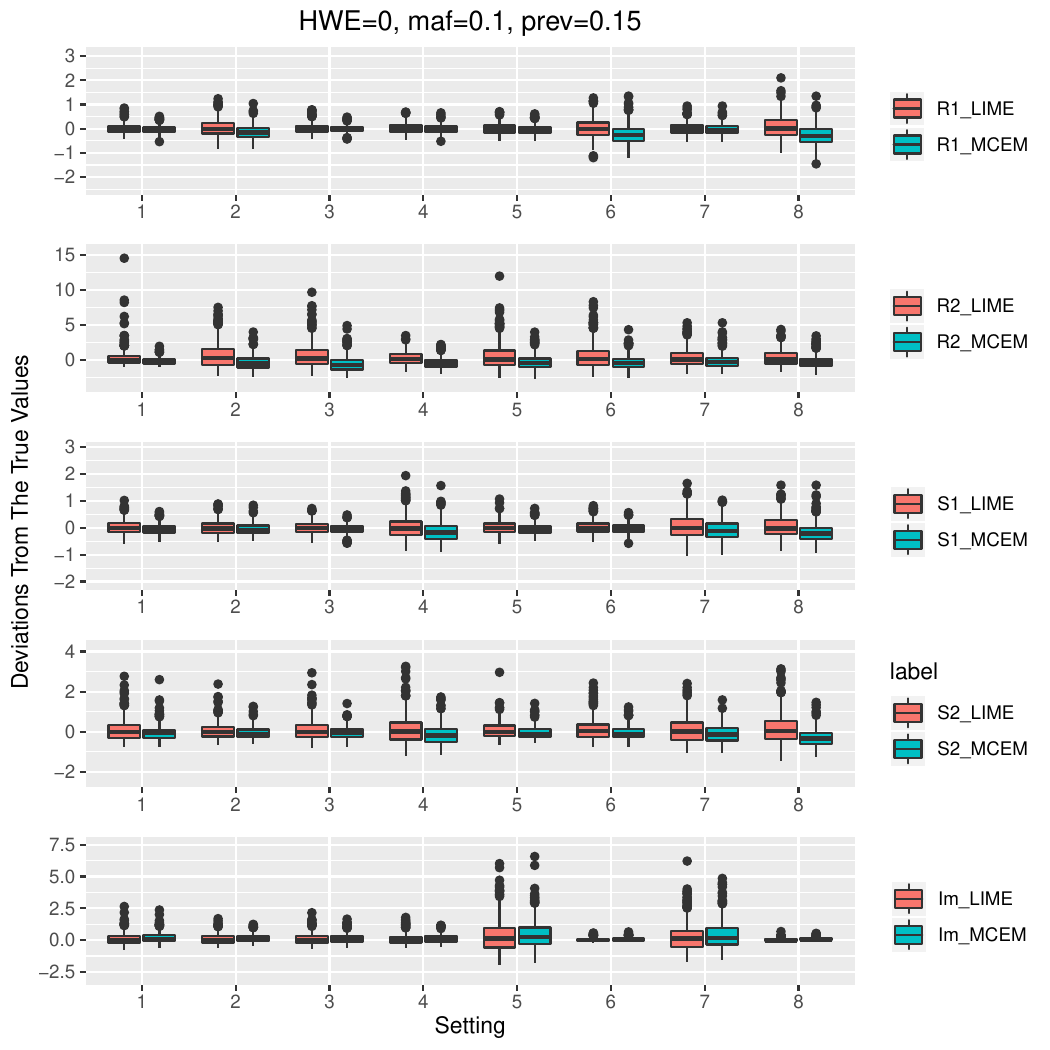}
\label{fig:subfig1}}
\qquad
\subfloat[Subfigure 2 list of figures text][]{
\includegraphics[width=0.4\textwidth]{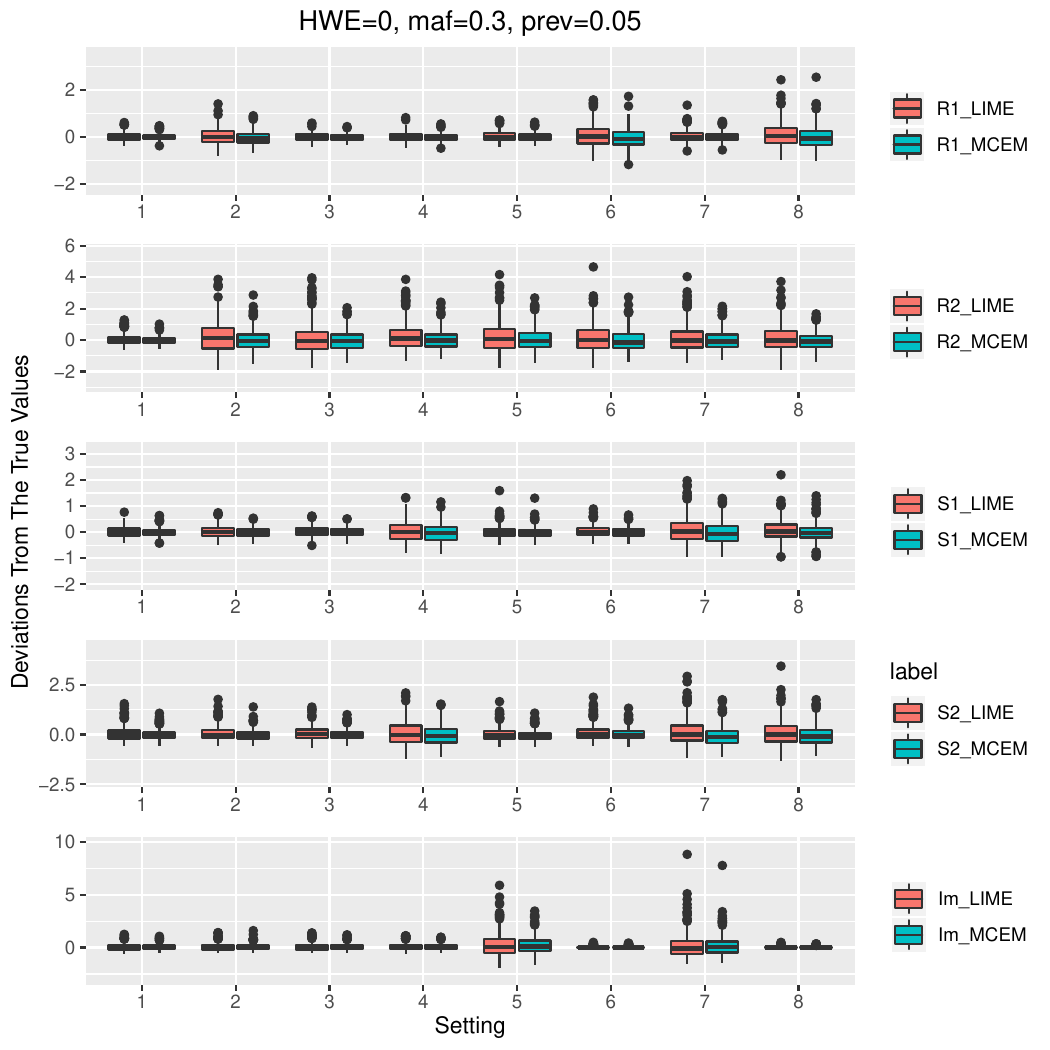}
\label{fig:subfig2}}
\qquad
\subfloat[Subfigure 3 list of figures text][]{
\includegraphics[width=0.4\textwidth]{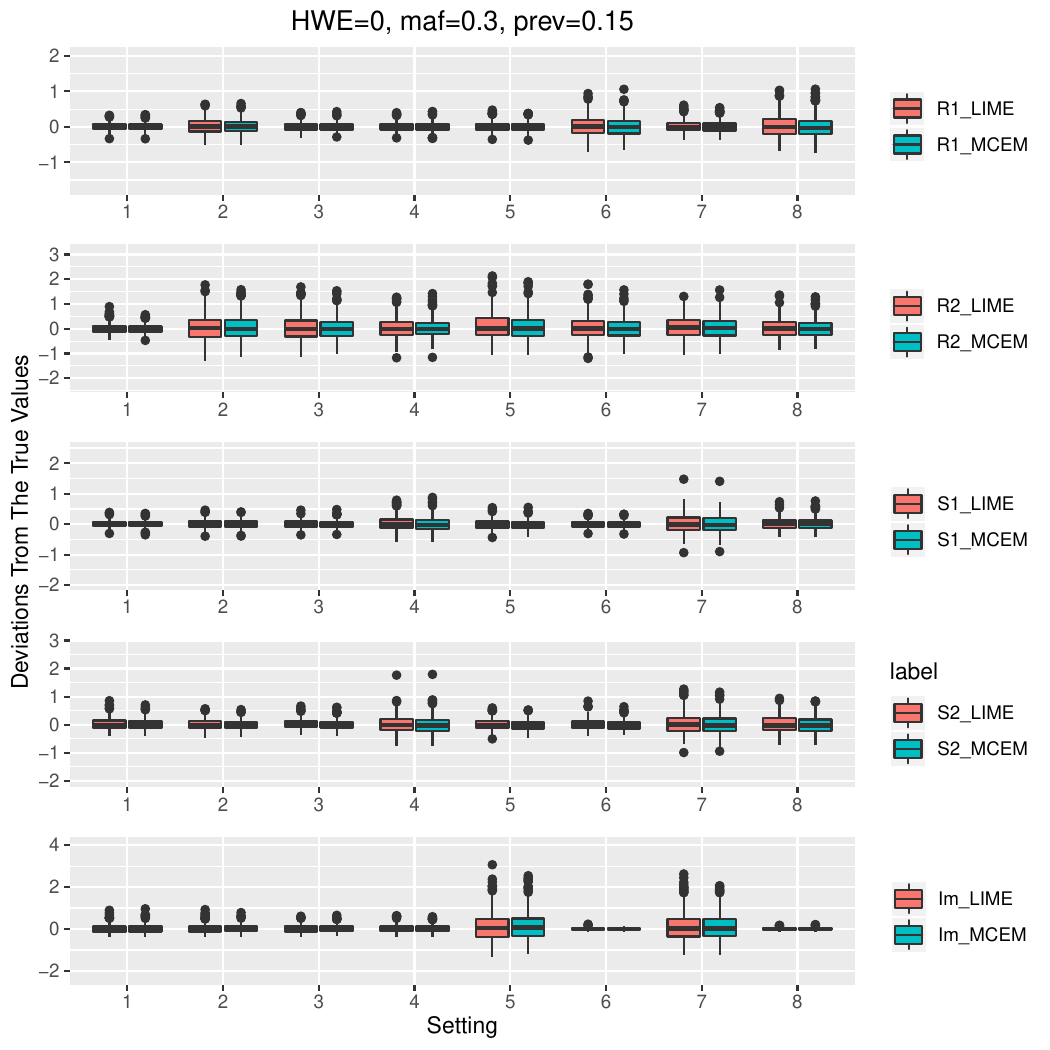}
\label{fig:subfig3}}
\qquad
\subfloat[Subfigure 4 list of figures text][]{
\includegraphics[width=0.4\textwidth]{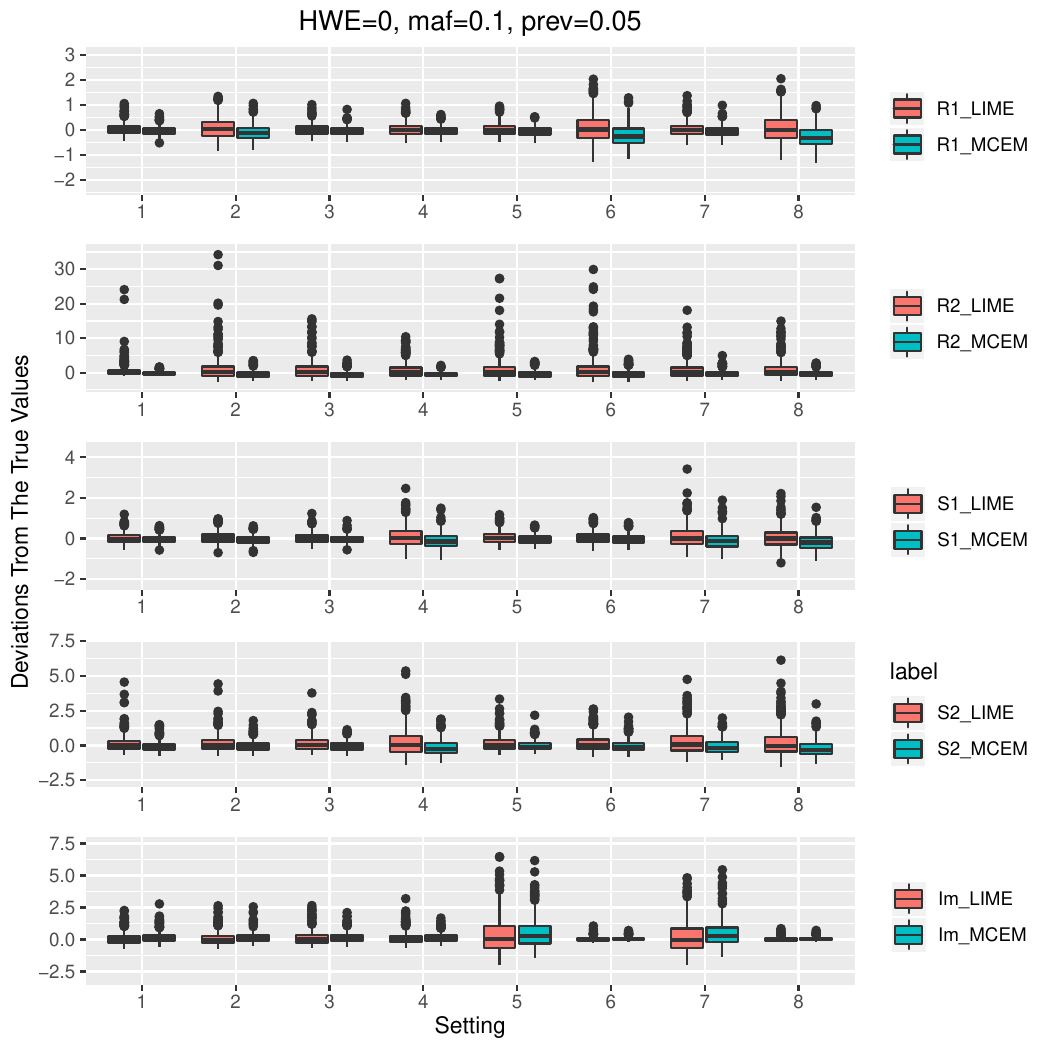}
\label{fig:subfig4}}
\caption{Box plots for biases of MCEM and LIME methods for the families with additional siblings where HWE=0.  }
\label{fig:globfig3}
\end{figure}

\clearpage
\begin{figure}[h]
\centering
\subfloat[Subfigure 5 list of figures text][]{
\includegraphics[width=0.4\textwidth]{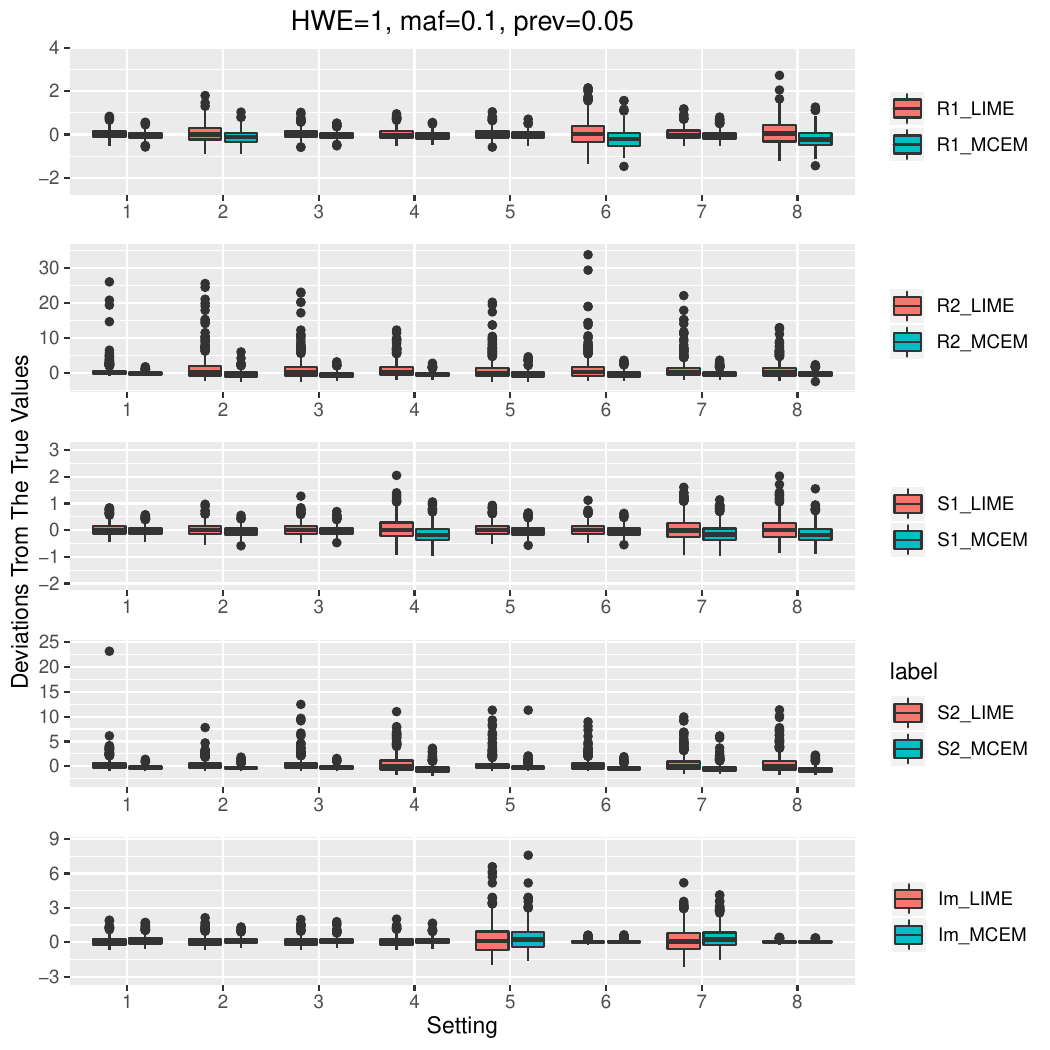}
\label{fig:subfig1}}
\qquad
\subfloat[Subfigure 6 list of figures text][]{
\includegraphics[width=0.4\textwidth]{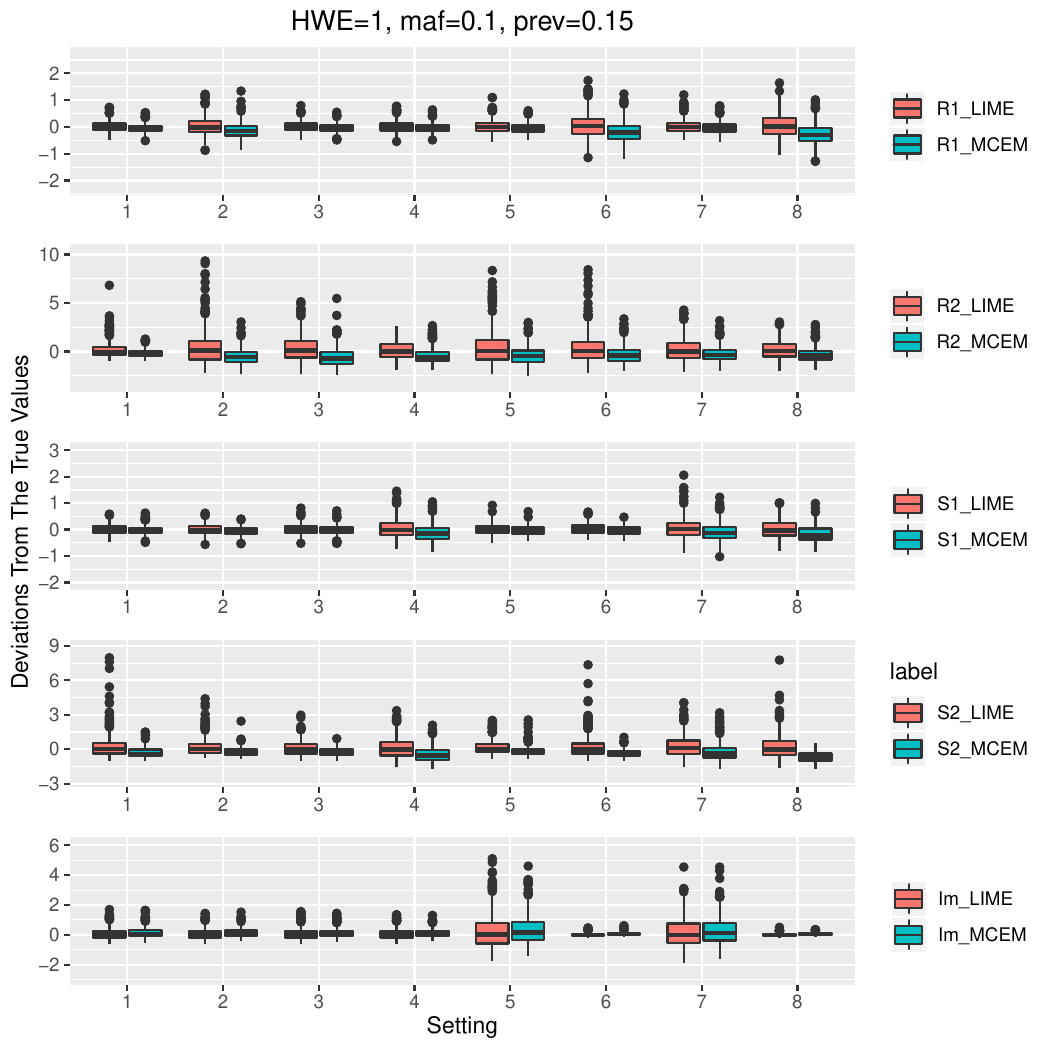}
\label{fig:subfig2}}
\qquad
\subfloat[Subfigure 7 list of figures text][]{
\includegraphics[width=0.4\textwidth]{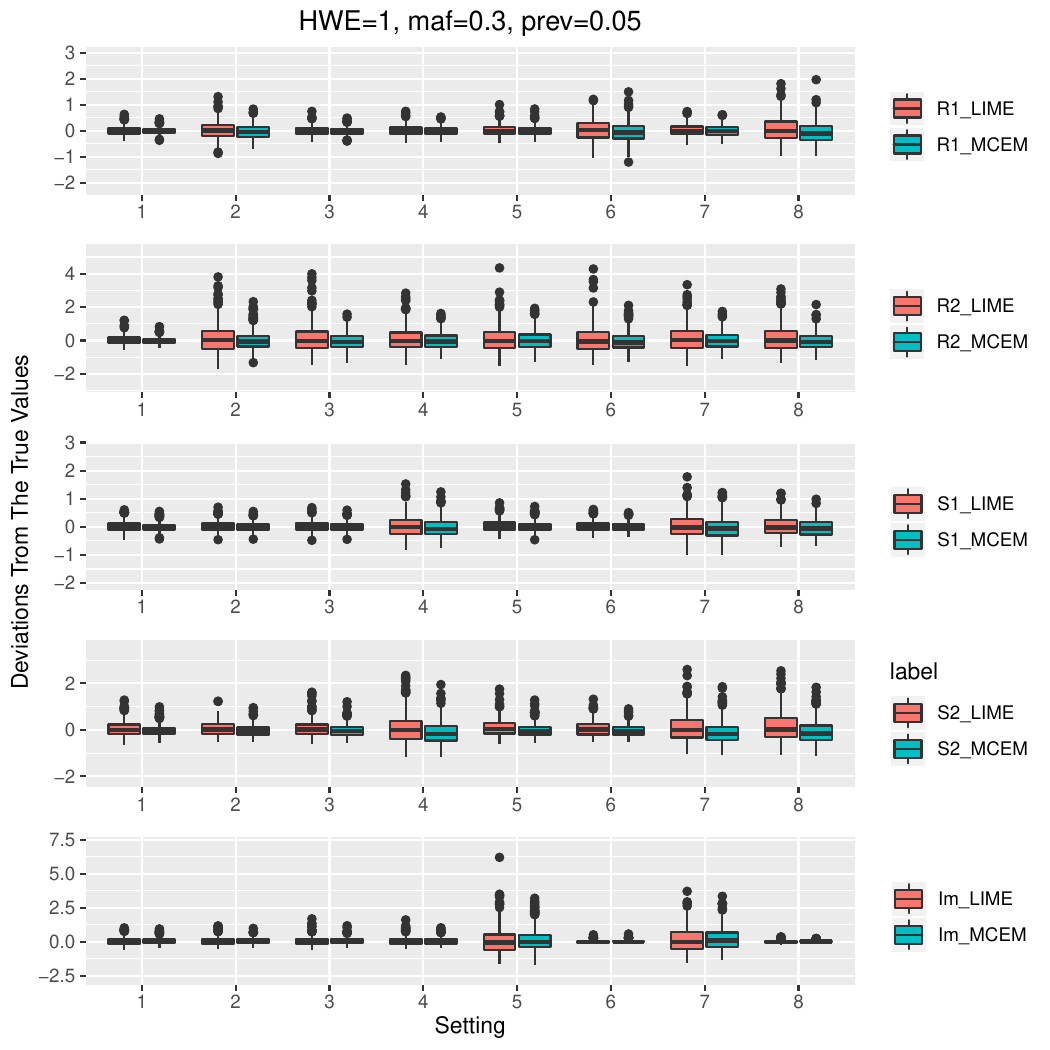}
\label{fig:subfig3}}
\qquad
\subfloat[Subfigure 8 list of figures text][]{
\includegraphics[width=0.4\textwidth]{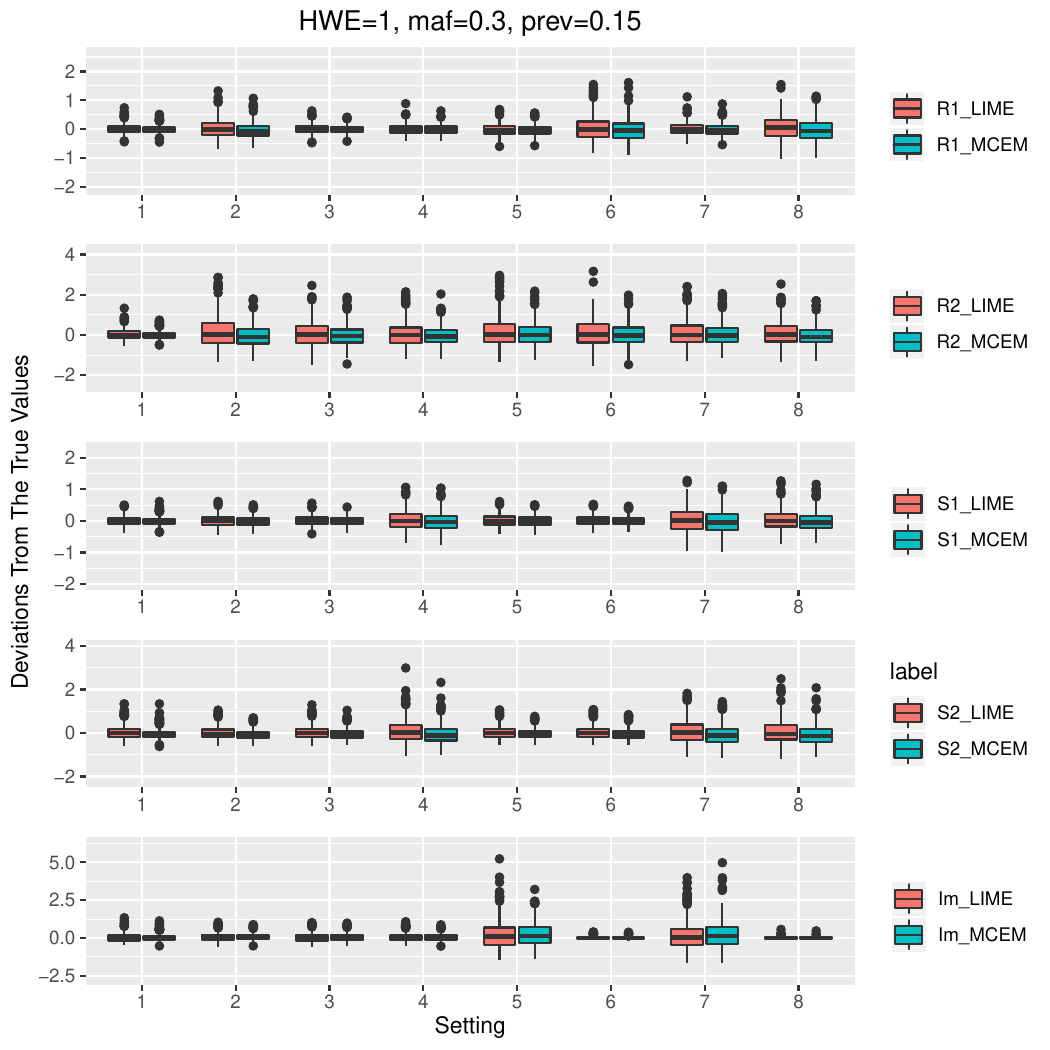}
\label{fig:subfig4}}
\caption{Box plots for biases of MCEM and LIME methods for the families with additional siblings where HWE=1.}
\label{fig:globfig4}
\end{figure}

\clearpage
\newpage

\begin{figure}[h]
\centering
\subfloat[Subfigure 5 list of figures text][]{
\includegraphics[width=0.4\textwidth]{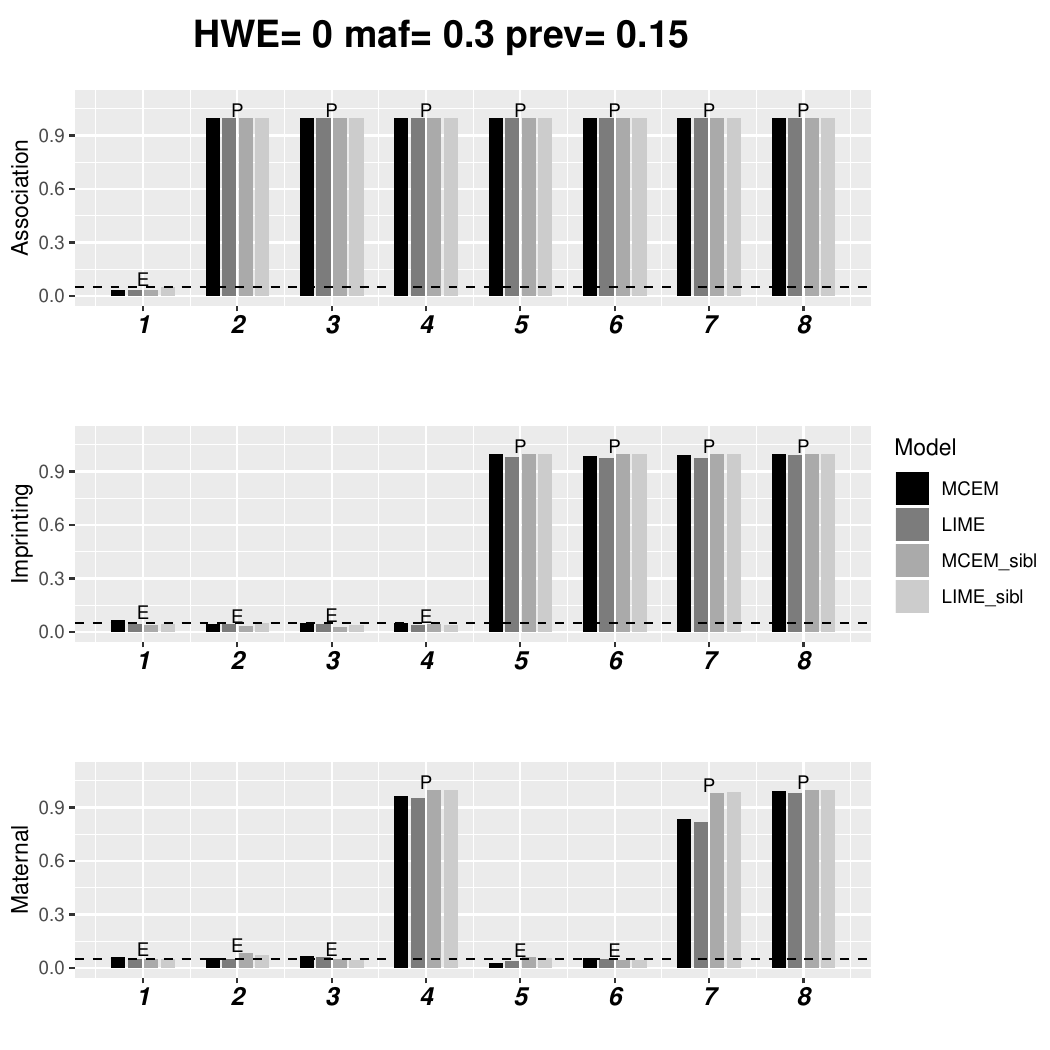}
\label{fig:subfig1}}
\qquad
\subfloat[Subfigure 6 list of figures text][]{
\includegraphics[width=0.4\textwidth]{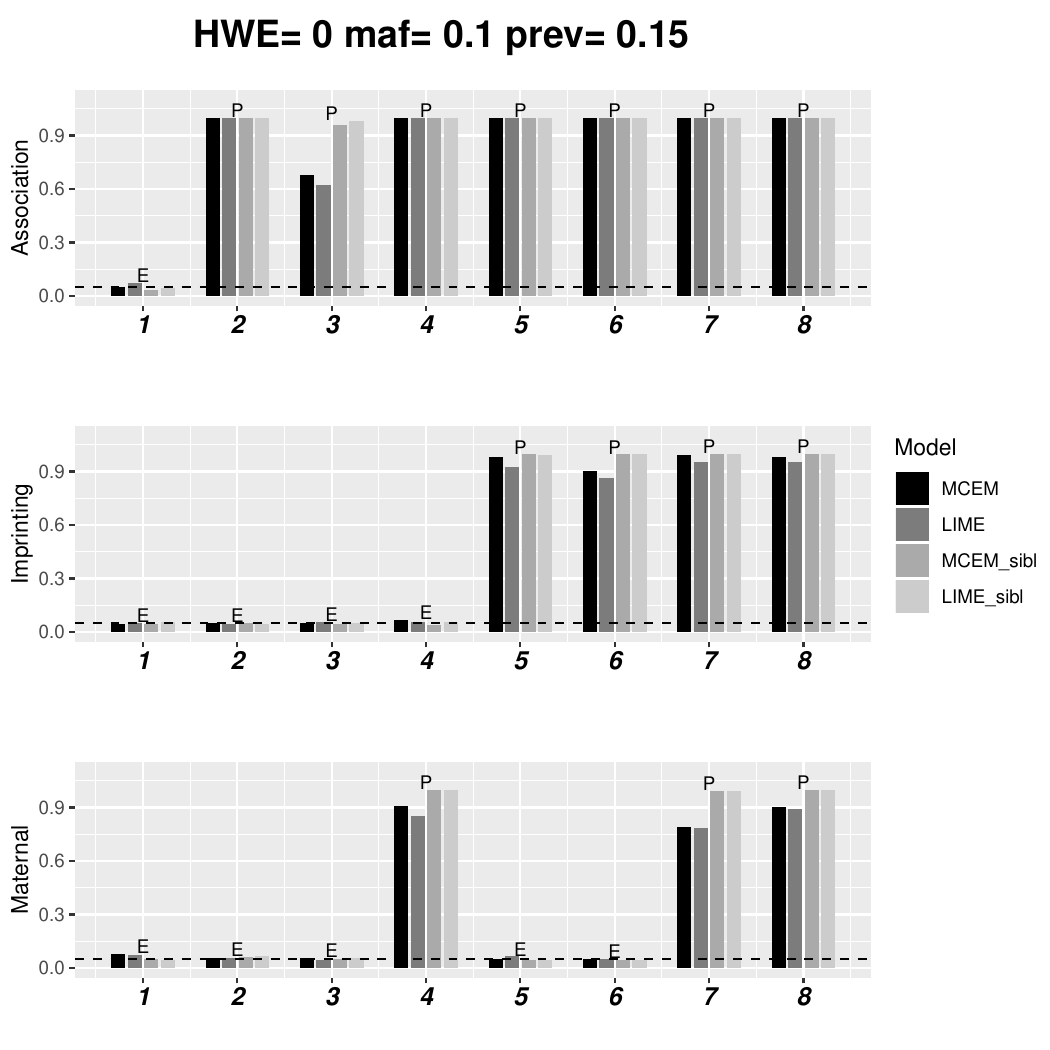}
\label{fig:subfig2}}
\qquad
\subfloat[Subfigure 7 list of figures text][]{
\includegraphics[width=0.4\textwidth]{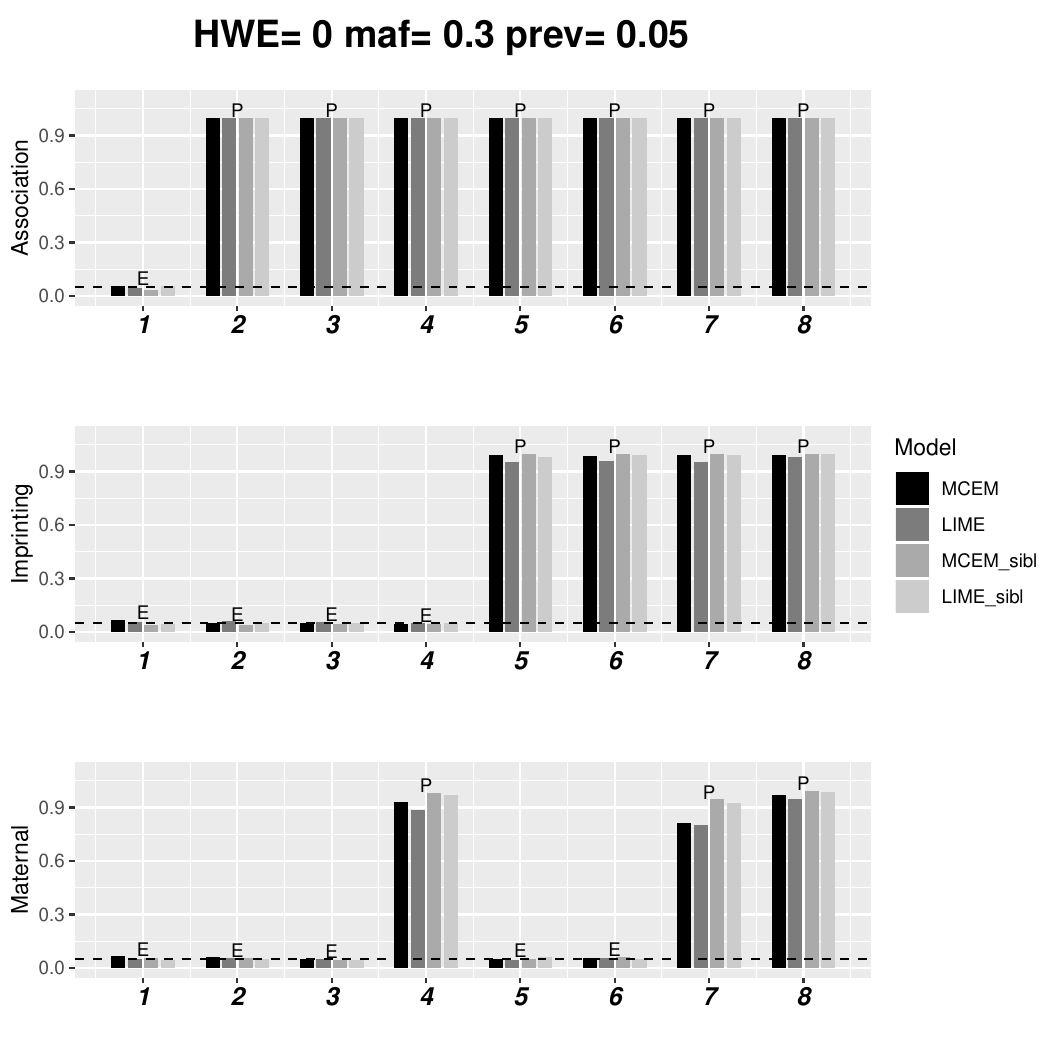}
\label{fig:subfig3}}
\qquad
\subfloat[Subfigure 8 list of figures text][]{
\includegraphics[width=0.4\textwidth]{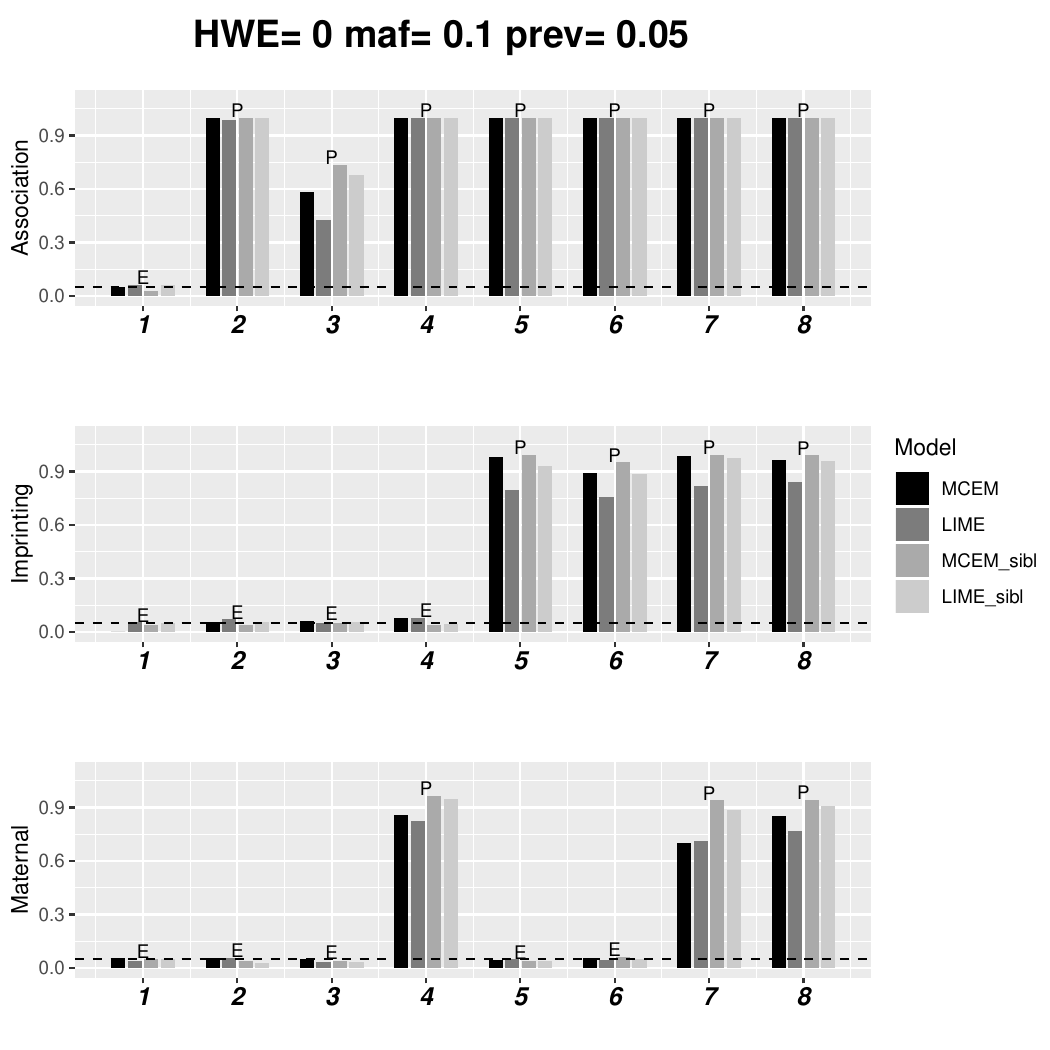}
\label{fig:subfig4}}
\caption{Barplots for type 1 error under HWE=0.}
\label{fig:globfig4}
\end{figure}

\clearpage

\bibliographystyle{apa}
\bibliography{citations}

\end{document}